\documentclass[fleqn,usenatbib]{mnras}

\usepackage{newtxtext,newtxmath}

\usepackage[T1]{fontenc}

\DeclareRobustCommand{\VAN}[3]{#2}
\let\VANthebibliography\thebibliography
\def\thebibliography{\DeclareRobustCommand{\VAN}[3]{##3}\VANthebibliography}

\usepackage{graphicx}	
\usepackage{amsmath}	

\title[Runaway collisions in nuclear stellar clusters]{Global instability by runaway collisions in nuclear stellar clusters: Numerical tests of a  route for massive black hole formation.}

\author[M.C. Vergara et al.]{M.C. Vergara$^{1,2}$\thanks{E-mail: marccortes@udec.cl (MCV)}, A. Escala$^{3}$, D.R.G. Schleicher$^{1}$ and B. Reinoso$^{4}$
\\
$^{1}$ Departamento de Astronom\'ia, Facultad Ciencias F\'isicas y Matem\'aticas, Universidad de Concepcion, Av. Esteban Iturra s/n Barrio Universitario,\\ Casilla 160-C, Concepcion, Chile
\\$^{2}$ Astronomisches Rechen-Institut, Zentrum für Astronomie, University of Heidelberg, Mönchhofstrasse 12-14, 69120, Heidelberg, Germany
\\$^{3}$ Departamento de Astronomía, Universidad de Chile, Casilla 36-D, Santiago, Chile
\\$^{4}$ Universität Heidelberg, Zentrum für Astronomie, Institut für Theoretische Astrophysik, Albert-Ueberle-Str. 2, 69120 Heidelberg, Germany
}

\date{Accepted XXX. Received YYY; in original form ZZZ}

\pubyear{2023}

\begin{document}
\label{firstpage}
\pagerange{\pageref{firstpage}--\pageref{lastpage}}
\maketitle

\begin{abstract}
The centers of galaxies host nuclear stellar clusters, supermassive black holes, or both. The origin of this dichotomy is still a mystery. Nuclear stellar clusters are the densest stellar system in the Universe, so they are ideal places for runaway collisions to occur. Previous studies have proposed the possible existence of a critical mass scale in such clusters, for which the occurrence of collisions becomes very frequent and leads to the formation of a very massive object. While it is difficult to directly probe this scenario with simulations, we here aim for a proof of concept using toy models where the occurrence of such a transition is shown based on simplified compact systems, where the typical evolution timescales will be faster compared to the real Universe. Indeed our simulations confirm that such a transition takes place and that up to 50\% of the cluster mass can go into the formation of a central massive object for clusters that are above the critical mass scale. Our results thus support the proposed new scenario on the basis of idealized simulations. A preliminary analysis of observed nuclear star clusters shows similar trends related to the critical mass as in our simulations. We further discuss the caveats for the application of the proposed scenario in real nuclear star clusters.
\end{abstract}

\begin{keywords}
methods: numerical – stars: kinematics and dynamics – black hole
\end{keywords}



\section{Introduction}

There is a dichotomy at the center of most galaxies, with some galaxies hosting supermassive black holes (SMBHs) at the center \citep{Volonteri2010, Kormendy2013}, and other galaxies harboring a Nuclear Star Cluster (NSC)  \citep{Boker2002, Cote2006}. In addition, the presence of NSCs surrounding SMBHs is also observed.
Any SMBH or/and NSC in the galaxy center is usually called a Central Massive Object (CMO) \citep{Ferrarese2006, Georgiev2016, Neumayer2020}. Nuclear star clusters are the densest stellar configurations in the Universe with masses of $\sim 10^6-10^{7}\rm~ M_\odot$ \citep{Walcher2005}, while supermassive black holes are one of the densest objects in the Universe with masses of $\sim 10^6-10^{10}\rm~ M_\odot$ \citep{Volonteri2010, Natarajan2009, King2016, Pacucci2017}. Additionally, there is evidence of a correlation between the mass of the SMBH and/or NSC with the properties in the host galaxy \citep{Ferrarese2006, Wehner2006, Li2007, Graham2009, Genzel2010, Leigh2012, Antonini2015}. A correlation between the mass of a SMBH and the velocity dispersion of the stars around it has been observed \citep{Ferrarese2000, Tremaine2002, Gultekin2009}, as well as a correlation between the mass of the SMBH and the mass of the bulge of its host galaxy \citep{Magorrian1998, Marconi2003, Haring2004}. On the other hand, there are also measurements of a correlation between the mass of the NSC and the bulge luminosity of their host galaxies \citep{Wehner2006, Cote2006, Rossa2006, Boker2008}, suggesting a possible  co-evolution between the SMBH and the NSC with the host galaxy. 

Recently the first image confirming the presence of the SMBH called Sagittarius A* in our Galaxy, the Milky Way \citep{EHT_sgtrA_a}, as well as studies of stellar orbits around this object \citep{Ghez2008, Genzel2010, Gillessen2017}, confirm that our galaxy has a NSC and a SMBH at its center \citep{Genzel2010, Schodel2014}.  Other galaxies with NSC and SMBH are e.g, NGC  4395 \citep{Filippenko2003}, M31\citep{Bender2005}, NGC  1042 \citep{Shields2008}, NGC  3621 \citep{Barth2009}, NGC  6388 \citep{Lutzgendorf2011}, NGC  404 \citep{Seth2010, Nguyen2017}, NGC  3319 \citep{Ning2018} and NGC  3593 \citep{Nguyen2022}. Observing galaxies that have NSCs but lack SMBHs poses a challenge due to the limitations of direct observation. One example is M33, which does not show any signs of an SMBH at its center \citep{Gebhardt2001}, in other galaxies such as NGC  300 and NGC  428 an SMBH has not been detected, but upper limits for their BH mass can be derived \citep{Neumayer2012}. The search for these SMBHs is a difficult task, resulting in only a few galaxies that provide useful measurements. Galaxies that have SMBHs but no NSCs are more frequently observed, especially for those with masses over $10^{11}\rm~ M_\odot$, as the presence of SMBHs often disrupts the formation of NSC \citep{Antonini2015}, such as NGC  4649 and NGC  4374 \citep{Neumayer2012}.

Moreover, there are SMBH detections at high redshift, though it is not yet clear how they could reach large masses of $\sim 10^6-10^{10}\rm~ M_\odot$ \citep{Volonteri2010, Natarajan2009, King2016, Pacucci2017} within such a short time, when the Universe was just a few hundred million years old. The high-redshift SMBH observations include e.g. the detection of the first  $3$ quasars at $z>6$ by \citet{Fan2003}, the most massive SMBH discovered by \citet{Wu2015} with a mass of $1.2\times10^{10}\rm~ M_\odot$ at $z=6.3$, and the most distant AGN at $z=7.642$ with a mass of $(1.6\pm 0.4)\times10^{9}\rm~ M_\odot$ observed by \citet{Wang2021}. In total more than $100$ quasars have been observed at $5.6\lesssim z\lesssim6.7$ as summarized by \citet{Banados2016}. \citet{Mortlock2011} observed a quasar at $z=7.085$ with a mass of $2\times10^{9}\rm~ M_\odot$.

Although the exact process of SMBH formation is still unknown, there are several scenarios that try to explain their formation \citep{Woods2019}, such as Direct Collapse (DC) based on the presence of massive gas clouds at high redshift, which collapse in an atomic cooling halo \citep{Bromm2003, Volonteri2008, Latif2013, Latif2015}. However, the presence of molecular hydrogen produces fragmentation in the cloud \citep{Omukai2008, Latif2016, Bovino2016, Suazo2019}. Forming a supermassive star requires 
accretion rates of $0.1~ \rm~ M_ {\odot}/yr$ \citep{Begelman2010, Schleicher2013, Sakurai2015}. Another scenario is based on the remnants of population III (Pop. III) stars; these stars are born in clouds with zero metallicity when the cloud collapses. The protostars accumulate material on the hydrostatic core, forming a very massive star (VMS) \citep{Omukai1998, Volonteri2003, Tan2004, Ricarte2018}. Another proposed scenario related to the stellar dynamics in a star cluster is runaway collisions and mergers. In this scenario, several collisions occur with a single star that eventually becomes a more massive object \citep{Rees1984, Devecchi2009, Katz2015, Sakurai2017, Sakurai2019, Boekholt2018, Reinoso2018, Reinoso2020, Alister2020, Vergara2021, Schleicher2022}. SMBH formation is possible through repeated mergers in dense star clusters. These kinds of mergers involve strong gravitational waves emission, which can be detected using current interferometers like LIGO\footnote{LIGO: \url{ https://www.ligo.caltech.edu/page/ligos-ifo}}, Virgo\footnote{Virgo: \url{https://www.virgo-gw.eu/}}, or Kagra\footnote{Kagra: \url{https://gwcenter.icrr.u-tokyo.ac.jp/en/}}, and in the future, with LISA\footnote{LISA: \url{https://lisa.nasa.gov}} and ET\footnote{ET: \url{https://www.et-gw.eu/}}
\citep{Fragione2020, Fragione2022}.

SMBHs are located in the galactic center. All surrounding gaseous material and stars will eventually lose their angular momentum as they fall into the center of the galaxy because it is the deepest gravitational zone of the stellar configuration \citep{Shlosman1990, Escala2006}. There are multiple processes that could lead to a strong inflow, e.g. gravitational torques during the merger of galaxies \citep{Barnes2002, Mayer2010, Prieto2021} or migration of groups by dynamical friction \citep{Escala2007, Elmegreen2008}, among others. At high redshift, there is a high fraction of systems with gas and no SMBH (which could provide AGN feedback), suggesting that the strong inflow should be more extreme \citep{Prieto2016}. Therefore, the densest stellar/gas configuration should be found at these locations. If these materials (gas and stars) do not form a SMBH, they will probably form NSCs, which is the other stable physical configuration that can remain in the center of the galaxy \citep{Escala2021}. NSCs are ideal places for stellar collisions to occur. Numerical simulations of clusters with masses less than $10^6 \rm~ M_\odot$ show a low black hole formation efficiency ($\epsilon_{BH}$) of up to a few percent of the initial mass \citep{Portegies2002, Devecchi2009, Sakurai2017, Reinoso2018}, while $\epsilon_{BH}$ grows dramatically for the most massive clusters ($10^7 \rm~ M_\odot$). This growth is triggered by runaways stellar collisions \citep{Lee1987,Quinlan1990,Davies2011,Stone2017}.
As SMBHs are significantly larger than MBHs, we will use the term MBHs going forward to maintain consistency.

In this paper, we explore a new scenario proposed by \citet{Escala2021} where stellar collisions in NSCs provide a mechanism to form MBHs. While the modeling of real nuclear star clusters will not be feasible, we will instead focus on simplified models where the relation between collision time, relaxation time and the available evolution time of the system can be expected to be self-similar. In section 2, we will introduce the model and the corresponding critical mass scale where collisions can be expected to become very efficient. Our approach to perform simulations where the existence of such a critical mass scale can be tested is then introduced in section 3. In section 4 we present our results. We discuss the observational counterpart of our simulations in section 5 and finally, we discuss them and our conclusions in section 6, where we also present the main caveats concerning the application of our results in real nuclear star clusters.

\section{Proposed scenario for MBH formation} \label{scenario}

\citet{Escala2021} proposed a new MBH formation scenario, motivated by  showing that the observed nuclear stellar clusters are in a regime where collisions are not relevant throughout the whole system; on the other hand, well-resolved observed MBHs are found in regimes where collisions are expected to be dynamically relevant. In this context, it was shown that NSCs in virial equilibrium with masses higher than $10^8\rm~ M_\odot$  have short collision times, so these stellar configurations are too dense to be globally stable against collisions. This leads to a destabilization of the cluster and allows most of the mass to collapse into a central massive object. 

NSCs are the densest stellar systems in nature, thus they represent one of the most favorable places for runaway collisions to occur. The close encounters of the stars within the cluster generally occur at high speed in the center due to the deep gravitational potential. The gravitational interactions can lead to the ejection of  stars that leave the system with some kinetic energy producing a redistribution of the energy and allowing the cluster to undergo a core-collapse \citep{Lynden1968, Cohn1979, Spitzer1987}. Stellar collisions are expected to occur when the cluster core collapses, causing a single object to experience almost all of the collisions, increasing exponentially in mass during the core collapse \citep{Portegies1999, Portegies2002}. The core-collapse has an associated time known as the relaxation timescale. If the system is virialized, the crossing time is $t_{cross}=\sqrt{R^{3}/GM}$, where G is the gravitational constant, and $R$ and $M$ are the radius and mass of the cluster, respectively. The crossing time quantifies the time that a star with a typical velocity needs to cross the cluster. The typical velocity is defined as the root mean square of the stellar velocities. The relaxation timescale is related to the perturbation of the global properties of the cluster such as stellar orbits and quantifies the energy exchange between two bodies. \citet{Binney2008} defined the relaxation timescale as $t_{relax}=0.1N*t_{cross}/\ln{(N)}$, where N is the total number of particles. 

In a cluster that has equal mass stars, the total number of stars is  $N=M/M_*$, where $M_*$ represents the mass of a single star, while the stellar radius is denoted as $R_*$. The occurrence of runaway collisions can be quantified through the collision timescale defined as  $t_{coll}=\lambda/\sigma$, where $\sigma$ is the velocity dispersion and $\lambda$ is the stellar mean free path \citep{Binney2008}. In a virialized system the velocity dispersion is defined as $\sigma=\sqrt{GM/R}$. \citet{Landau1980} and \citet{Shu1991} define a probabilistic mean free path as $\lambda=1/n\Sigma_0$, where $\Sigma_0$ is the effective cross-section and $n$ the number density of stars. Therefore the collision rate is defined as $t_{coll}=\sqrt{R/GM(n\Sigma_0)^2}$, the number density is $n=3M/4\pi R^3 M_*$ and the effective cross-section is $\Sigma_0 = 16 \sqrt{\pi} R_*^2(1+ \Theta)$, where $\Theta=9.54 ((M_* R_\odot)/(M_\odot R_*))(\rm~ (100\, kms^{-1})/(\sigma))^2$ is the Safronov number \citep{Binney2008}.
Under the condition that the collision time is equal to or shorter than the age of the system $(t_H)$, we can derive the following equation,

\begin{equation} 
M\geq \left(\frac{4 \, \pi \, M_*}{3 \, \Sigma_0 \, t_H \, G^{1/2}}\right)^{2/3}R^{7/3},
\label{coll_ts}
\end{equation}
and under the condition that the relaxation time is equal to or shorter than the time $t_H$, we have
\begin{equation}
R\leq\left(\frac{t_H\, M_*}{0.1}\ln\left(\frac{ M}{M_*}\right)\right)^{2/3}\left(\frac{G}{M}\right)^{1/3} .
\label{relax_ts}
\end{equation}

Note that for $\rm~ M_*= 1\rm~ M_\odot$, $\rm~ R_*= 1\rm~ R_\odot$ and $\sigma=100\rm~ km/s$, the Safronov number is $\Theta=9.54$, leaving a value for the effective cross-section of $\Sigma_0\approx100 \pi \rm~ R_\odot^2$. Equations~\ref{coll_ts} \& \ref{relax_ts} lead to the conditions discussed in \citet{Escala2021}\footnote{We note a typo in Eq.~2 of \citet{Escala2021} introduced by the journal during the 
proofreading process: $\rm~ t_{HG}^{1/2}$ in \citet{Escala2021} should be $\rm~ t_{H}G^{1/2}$} in their equations 2 \& 4. 

The dichotomy in the centers of galaxies shows CMOs, which can be MBHs and/or NSCs. The presence of NSCs is generally found in galaxies with masses less than $10^{10}\rm~ M_\odot$, while MBHs are generally found in more massive galaxies with masses larger than $10^{12}\rm~ M_\odot$. There is also evidence of the coexistence of both objects in galaxies with intermediate masses \citep{Georgiev2016}. It is possible that both objects are in a different phase of evolution from a common formation mechanism. If the CMO is too dense and meets the condition of having a collision time shorter than the time $t_H$, then the system cannot expand and becomes globally unstable against collisions, and probably collapses into a MBH. On the other hand, for less dense CMOs, the system is globally stable, allowing the presence of an NSC that probably coexists with a low-mass BH in the center \citep{Escala2021}.

Simulations of stellar dynamics have been investigated using N-body codes for cluster masses typically smaller than $10^6 \rm~ M_\odot$ \citep{Portegies2002, Devecchi2009, Sakurai2017, Reinoso2018}. Since more massive systems are numerically expensive, these configurations have been investigated with Fokker-Planck models of galactic nuclei developed by \citet{Lee1987, Quinlan1990}, which show a transition called 'merger instability' in the formation of the CMO at masses over or similar to $10^7\rm~ M_\odot$. However, in order to test the global instability proposed by \citet{Escala2021}, it  is useful to define a critical mass ($M_{crit}$) as the mass at which the virial radius of our clusters crosses the collision line in Fig.~\ref{ic_10myr}. This can be computed from the marginal condition in Eq.~\ref{coll_ts}:

\begin{equation} \label{eq_mass_crit}
    M_{crit}=R^{7/3}\left(\frac{4\,\pi \,M_*}{3\,\Sigma_0\,t_H\,G^{1/2}}\right)^{2/3}.
\end{equation}

In the next section, we will describe the initial conditions of our simulations. These are meant to provide a proof of concept for the new scenario, showing that a transition exists at the critical mass derived by \citet{Escala2021}. We note here in advance that for computational reasons, these simulations are pursued for less massive but more compact clusters considering also a shorter evolution time, as in that case, the critical mass scale is lowered and it becomes feasible to explore the transition via N-body simulations. In the real Universe, clusters are less compact but have much more time to evolve, leading to critical masses of the order $10^7\rm~ M_\odot$. 

\section{Model setup and simulations} \label{simulations}

Our simulations were run with \href{https://github.com/nbodyx/Nbody6ppGPU}{{\sc nbody6++gpu}} \citep{Wang2015}, a direct N-body code with high precision based on the codes  {\sc nbody6++} \citep{Spurzem1999} and {\sc nbody6} \citep{Aarseth2000}. This code includes an algorithm to solve N-body interactions such as close encounters, binaries \citep{Kustaanheimo}, or multiple systems \citep{Mikkola1990, Mikkola1993}. Besides,  it includes a spatial hierarchy to speed up computational calculations \citep{Ahmad1973}. This code works with the $4^{th}$ order Hermite integrator scheme of \citet{Makino1991} and also includes optimizations for the calculations of gravitational forces between particles using Graphics Processing Units (GPUs) \citep{Nitadori2012, Wang2015}.

We consider that collisions occur when the radii of two stars overlap (i.e $D \leq R_{*1} + R_{*2}$); if the stars fulfill this condition, we replace them with a new single star. We neglect mass loss here since is usually a small percentage (<$10\%$) \citep{Dale2006, Gaburov2008, Glebbeek2008, Alister2020}. Note that in \citet{Alister2020} despite the fact that it may seem insignificant, a loss of just $1\%$ in mass can become a significant factor in simulations involving collisions of large objects, such as Population III stars, which is why it remained relevant in their simulations. In the toy models presented here, we are not going to such high masses, for which eventually the mass loss may balance the mass gain during the evolution. The parameters of the new star after the merger are as follows:

\begin{equation}
M_{new}=M_{*1}+M_{*2},
\end{equation}

\begin{equation}
R_{new}=R_{*1}(1+M_{*2}/M_{*1})^{\frac{1}{3}},
\end{equation}

which assumes that the stellar mass density remains constant. We consider that the new star reaches hydrostatic and thermal equilibrium quickly after the collision.

For more details about computational accuracy, check the appendix~\ref{App1}

\subsection{Initial Conditions}

We are testing the new scenario of NSC instability under collisions as a mechanism to form MBHs \citep{Escala2021}, using a \citet{Plummer1911} distribution of equal-mass stars and evolving the clusters for a time $t_H=10\,$Myr. We also vary the initial number of stars as $N=5\times10^2, 10^3, 5\times10^3, 10^4$ and the initial cluster virial radii as $R_v=0.005, 0.01, 0.1, 1\rm~pc$.  The stars have an initial mass of $M_* = 1, 10, 50\rm~M_\odot$ with a radius  $R_* = 1, 4.7, 11.7\rm~R_\odot$,  respectively, computed from the  mass-radius relation

\begin{equation}
R_* = 1.6\times\left(M_*/\rm~ M_\odot\right)^{0.47}\rm~ R_\odot, \hspace{1.6cm} 10\leq M_*< 50\rm~ M_\odot,
\end{equation}

\begin{equation} 
R_*=0.85\times\left(M_*/\rm~ M_\odot\right)^{0.67}\rm~ R_\odot, \hspace{1.6cm}M_* \geq  50\rm~ M_\odot,
\end{equation}
from the relations described by \citet{Bond1984} and \citet{Demircan1991}, respectively.

We chose these initial conditions with clusters that are less massive but denser than real ones due to computational limitations, as it would be unfeasible to run clusters of realistic sizes and masses for a sufficiently long evolutionary time. Instead we explore if the expected instability occurs with simplified simulations using more compact systems but also shorter timescales, comparing simulations above and below the expected critical mass scale that is calculated from the system parameters. These simulations thus provide a proof of concept to demonstrate if the expected transition occurs in a simplified system. 

\begin{figure}
    \centering    \includegraphics[width=\columnwidth,height=\columnwidth]{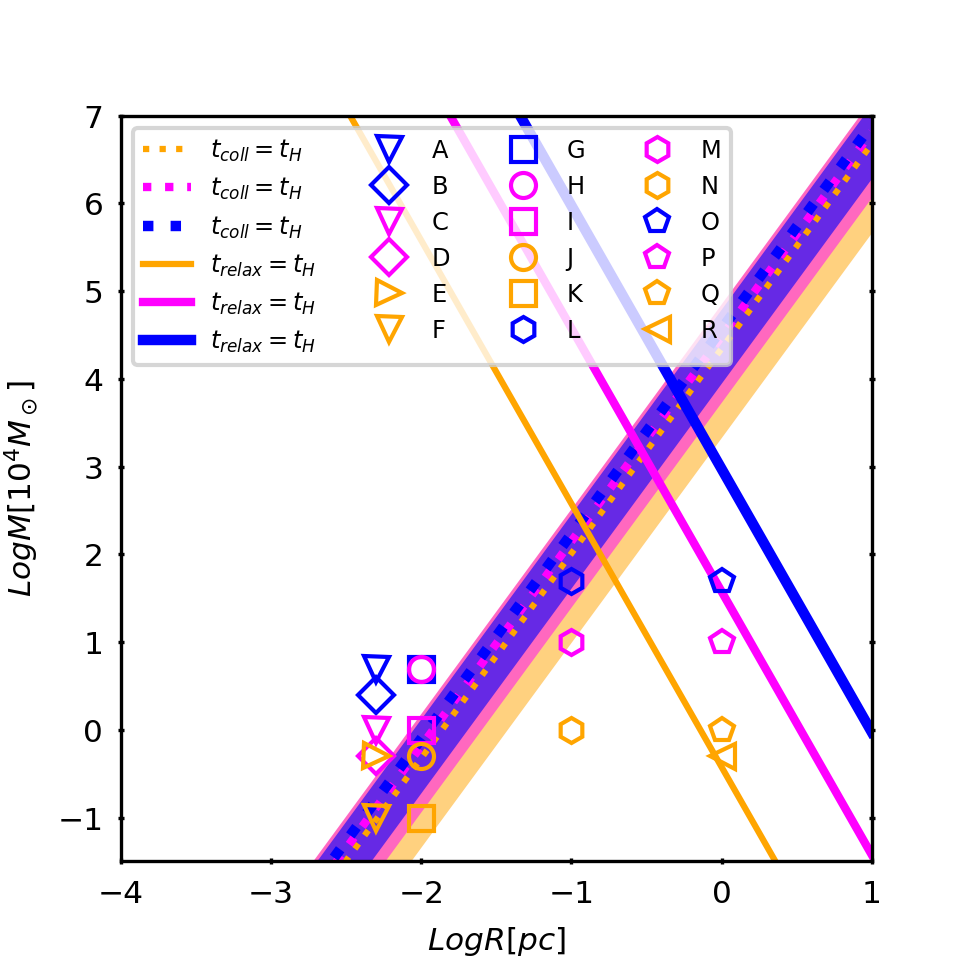}
    \caption{The blue, magenta and orange symbols, lines and colored areas are for $M_* = 50,10, 1\rm~M_\odot$, respectively. Equal symbols represent the same number of stars $N$. Solid lines are from the condition of Eq.~\ref{relax_ts}, dotted lines are the mean of Eq.~\ref{coll_ts} for the different models, and all lines for the time evolution of $t_H = 10\,$Myr. The colored areas represent the range of the condition given by Eq.~\ref{coll_ts}, for the  models with different initial $\sigma$ (Table~\ref{ic}).}
    \label{ic_10myr}
\end{figure}

In Fig.~\ref{ic_10myr} we show the distribution of our models in the mass-radius parameter space. The equations \ref{coll_ts} and \ref{relax_ts}, for $t_H=10\,$Myr are described by the orange lines  for $M_*=1\rm~M_\odot$,  magenta lines  for $M_*=10\rm~M_\odot$ and blue lines for $M_*=50\rm~M_\odot$ (dotted lines for Eq.~\ref{coll_ts} and solid lines for Eq.~\ref{relax_ts}). The hatched area shows the range of Eq.~\ref{coll_ts} for different initial velocity dispersion $(\sigma)$. The left side of the hatched area $(t_{coll} <t_{H})$ is the parameter space where collisions dominate the cluster dynamics over the whole system, while the right side $(t_{coll} >t_{H})$ represents the parameter space where collisions are not globally relevant. Nevertheless, in the core of the cluster the collisions occur more frequently; then the NSCs can coexist with an unstable core triggered by the Spitzer instability \citep{Spitzer1969, Portegies2002}. Therefore, the collision timescales are important to determine the critical mass at which  the global instability will occur in NSCs against runaway collisions. 

Observed  NSCs typically have  radii of $1-10 \rm~ pc$  and masses  around $10^6-10^8\rm~ M_\odot$ \citep{Georgiev2016}, still being  globally stable, since they  have a collision time larger than the age of the Universe \citep{Escala2021} (i.e. they are one order of magnitude in mass below the line given by Eq.~\ref{coll_ts}). The critical mass is typically in the range $10^7-10^9\rm~ M_\odot$. Performing simulations in this range of NSC masses is still  numerically prohibitive (for  $\rm 1-10 \, M_\odot$ stars). However, we can run simulations of less massive NSCs, which are numerically less expensive when run over a shorter evolutionary period. 

In particular, we present here a set of simulations that crosses the collision threshold given by Eq.~\ref{coll_ts} to test the catastrophic dynamics of the proposed scenario. We simulate NSCs above the line described by Eq.~\ref{coll_ts} (i.e. lower critical masses ($10^3-10^5\rm~ M_\odot$) ), using a radius range of $0.005-0.01\rm~ pc$ for a mass range of $10^3-10^4\rm~ M_\odot$. In addition, some of our models with a radius of $0.1-1 \rm~ pc$ have a mass from around $10^3-10^5\rm~ M_\odot$; this means that they are below the line described by Eq.~\ref{coll_ts}, having larger critical mass than cluster masses, in order to check how the efficiency of forming a MBH changes as we cross the line that defines the marginal condition in Eq.~\ref{coll_ts}.
 
 \begin{table}
    \centering
    \caption{The virial radius is $R_v$, $M$ is the initial mass of the cluster, $N$ is the initial number of stars, the stellar mass and radius is $M_*$ and $R_*$, respectively, and $\sigma$ is the velocity dispersion.}
    \begin{tabular}[m{0.5\textwidth}]{cllllll}\hline\hline
       Models & $R_V$ & $M$ & $N$  & $M_*$ & $R_*$ & $\sigma$\\
       ID& $[\mathrm{pc}]$ & $[\mathrm{M_\odot}]$ & & $[\mathrm{M_\odot}]$ & $[\mathrm{R_\odot}]$ & [km/s] \\\hline
       A & $0.005$ & $5\times 10^4 $ & $10^3 $ & $50$ & $11.7$ &$207.38$\\
       B & $0.005$ & $2.5\times 10^4 $ &$5\times 10^2$& $50$ & $11.7$ & $146.64$\\
       C & $0.005$ & $10^4 $ & $10^3 $ & $10$ & $4.7$ & $92.74$\\
       D & $0.005$ & $5\times 10^3 $ & $5\times10^2$ & $10$ & $4.7$ &$65.58$\\
       E & $0.005$ & $5\times 10^3 $ & $5\times 10^3$ & $1$ & $1$ &$65.58$\\
       F & $0.005$ & $10^3 $ & $10^3 $ & $1$ & $1$ &$29.32$\\
       G & $0.01$ & $5\times 10^4 $ & $10^3 $ & $50$ & $11.7$  &$146.64$ \\
       H & $0.01$ & $5\times 10^4 $ & $5\times 10^3$ & $10$ & $4.7$ &$146.64$\\
       I & $0.01$ & $10^4 $ & $10^3 $ & $10$ & $4.7$ &$65.58$\\
       J & $0.01$ & $5\times 10^3 $ & $5\times 10^3$ & $1$ & $1$ & $46.37$\\
       K & $0.01$ & $10^3 $ & $10^3 $ & $1$ & $1$ & $20.73$\\
       L & $0.1$ & $5\times 10^5 $ & $10^4 $ & $50$ & $11.7$  &$146.64$\\
       M & $0.1$ & $10^5 $ & $10^4 $ & $10$ & $4.7$ &$65.58$\\
       N & $0.1$ & $10^4 $ & $10^4 $ & $1$ & $1$ &$20.73$\\
       O & $1$ & $5\times 10^5 $ & $10^4 $ & $50$ & $11.7$ &  $46.37$\\
       P & $1$ & $10^5 $ & $10^4 $ & $10$ & $4.7$ & $20.73$\\
       Q & $1$ & $10^4 $ & $10^4 $ & $1$ & $1$ &$6.55$\\
       R & $1$ & $5\times 10^3 $ & $5\times 10^3$ & $1$ & $1$ &$4.63$
       \\
    \hline   
    \end{tabular}
    \label{ic}
\end{table}
 
 The idealized models presented here are useful for testing the concept described by \citet{Escala2021} and this is the first step in testing this new proposed scenario considering global instabilities in NSCs as a mechanism to form MBHs. Our models are more representative of when the NSCs are born, as the NSCs can have a radius approximately ten times smaller at the moment of formation \citep{Banerjee2017}. Due to the evolution process, it must be considered that the radius of the cluster expands \citep{Baumgardt2018, Panamarev2019}, moving the models from left to right.

We simulated 18 nuclear star clusters, covering different regions of the mass-radius parameter space. The empty blue symbols are for $M_*=50\rm~ M_\odot$, empty magenta symbols for $M_*=10\rm~ M_\odot$, and empty orange symbols for $M_*=1 \rm~ M_\odot$. It is important to note that our NSC models do not reach masses greater than $10^7\rm~ M_\odot$ (much less with $10^7$ particles). However, we can still explore this new scenario proposed by \citet{Escala2021} via the modeling of more compact clusters for shorter time intervals. For such clusters, the critical mass scale from Eq.~\ref{eq_mass_crit} will be reduced, so that the computational modeling becomes feasible due to the smaller number of stars in the simulation. We summarize our initial conditions in table~\ref{ic}.

It is very important to highlight that our toy models were chosen according to the initial conditions given by Fig.~\ref{ic_10myr}, which is related to Eq.~\ref{coll_ts} and \ref{relax_ts}, derived from \citet{Binney2008} definition of collision time and relaxation time, respectively. Also, it is important to note the limited computational resources that do not allow exploring this new scenario for larger masses and radii (i.e. more realistic NSCs). However, using these toy models as proof of concept is the first step towards a more realistic model to prove that any NSC that fulfills the conditions given in Fig.~\ref{ic_10myr} (i.e being above the dotted lines given by Eq.~\ref{coll_ts}), must show a similar behavior during its evolution.



\section{Results}

The initial mass of the clusters in our models is $\rm~ M= N\times M_*$. The mass of the CMO can be computed as the sum of the masses of the NSC and BH ($ M_{CMO} = M_{NSC} + M_{BH} $), the mass of the BH can be denoted by $M_{BH}=\epsilon_{BH}M_{CMO}$, and the mass of the surrounding stars of the NSC can be denoted as $M_{NSC}=(1-\epsilon_{BH})M_{CMO}$. Therefore, the black hole formation efficiency ($\epsilon_{BH}$) can be expressed  as

\begin{equation}
\epsilon_{BH}=(1+M_{NSC}/M_{BH})^{-1}.
\label{eff_bh}
\end{equation}

From the simulation data the mass of the CMO can be calculated as $M_{CMO}=M-M_{esc}$, where $M_{esc}$ is the cumulative mass of the ejected stars, thus

\begin{equation}
M=M_{CMO}+M_{esc}= M_{NSC} + M_{BH}+M_{esc}.
\end{equation}

\subsection{Simulated star cluster evolution}

In this subsection, we analyze the time evolution of the cumulative mass of stars that escape from the system, the growth of mass of the most massive object, the black hole formation efficiency, and the Lagrangian radii at $90\%$, $50\%$, and $10\%$.

\begin{figure}
    \centering
    \includegraphics[width=\linewidth]{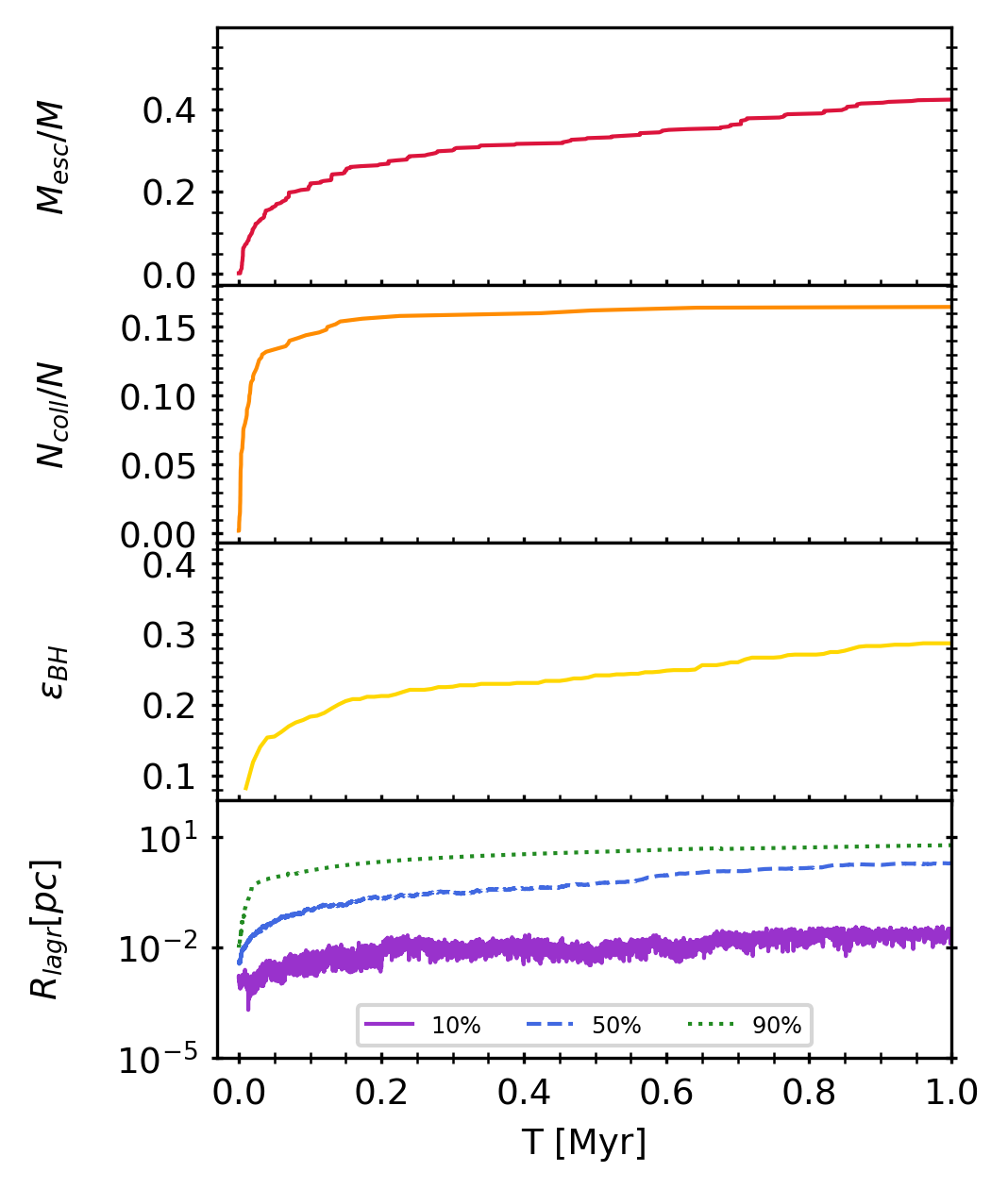} 
    \caption{Evolution of model B for $1\,$Myr. Top panel: The cumulative mass of escapers normalized by the initial mass $M$. First middle panel: The number of collisions normalized by the initial number of stars $N$. Second middle panel: The black hole formation efficiency $\epsilon_{BH}$ described by Eq.~\ref{eff_bh}. Bottom panel: Lagrangian radii for the $10\%$, $50\%$, and $90\%$ of the enclosed mass.}
    \label{C1Myr}
\end{figure}

\begin{figure}
    \centering
    \includegraphics[width=\linewidth]{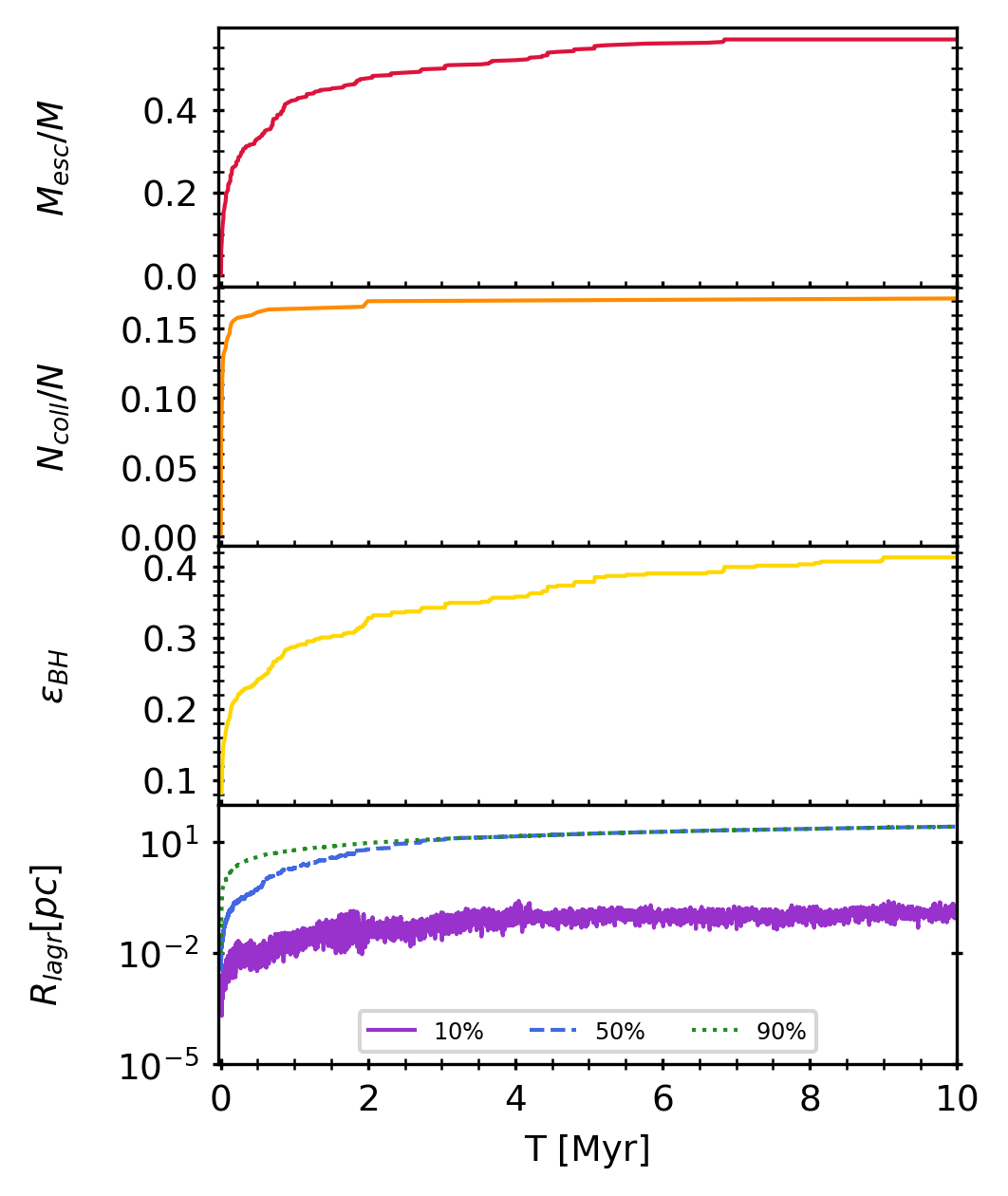} 
    \caption{Evolution of model B for a time period of $10\,$Myr. Panels are the same as in Fig.~\ref{C1Myr}.}
    \label{C10Myr}
\end{figure}

We analyze the evolution of two models, B and M. Model B has $M=2.5\times10^4\rm~ M_\odot$, with a virial radius of $0.005\rm~ pc$. Model M has an initial mass of $10^5\rm~ M_\odot$, with $R_v=1\rm~ pc$. Model B is in the region where collisions dominate the stellar dynamics (i.e. $t_{coll}<t_{H}$). Model M is in the region where collisions are not relevant over the whole system  (i.e. $t_{coll}>t_{H}$). We present the cumulative mass of escapers normalized by the initial mass $M$, the number of collisions normalized by the initial number of stars $N$, the black hole formation efficiency $\epsilon_{BH}$ described by Eq.~\ref{eff_bh} and the  Lagrangian radii corresponding to $90\%$, $50\%$, and $10\%$ of the enclosed mass.

In Fig.~\ref{C1Myr} we show the evolution of model B over $1\,$Myr. The top panel shows the cumulative mass of stars escaping from the cluster normalized by the initial mass $M$; the stellar cluster has lost around $42\%$ of the initial mass after $1\,$Myr. The first middle panel shows the total number of collisions $N_{coll}$ normalized by the initial number of stars $N=500$; until $1\,$Myr $82$ collisions have occurred. The second middle panel shows the black hole formation efficiency $\epsilon_{BH}$ reaching a value of around $28\%$. The stellar system forms a single massive object of $4150\rm~{M_\odot}$. The bottom panel shows the Lagrangian radii at $90\%$, $50\%$, and $10\%$ of the enclosed mass. The outer zone of the stellar system corresponding to $90\%$ of the mass shows an expansion until around $0.1\,$Myr, then it remains almost constant.
The middle zone corresponding to $50\%$ of the mass shows a smooth expansion all the time while the inner zone at $10\%$ of the enclosed mass shows a decrease at the beginning of the simulation.

In Fig.~\ref{C10Myr} we display the evolution of the same model B over $10\,$Myr. The panels are the same as in Fig.~\ref{C1Myr}. The top panel shows that around $58\%$ of the initial mass is lost. The first middle panel shows that $85$ collisions in total occurred, most collisions happened in the first Myr. The most massive object reaches a mass of $4300\rm~{M_\odot}$.
The second middle panel shows that the black hole formation efficiency increases until a value of $41\%$. After $2\,$Myr, the $90\%$ and $50\%$ Lagrangian radii curves start to overlap because there is so much mass loss. The $10\%$ Lagrangian radius shows a rapid decrease at first followed by an expansion and then remains almost constant.

Comparing Fig.~\ref{C1Myr} and Fig.~\ref{C10Myr}, after $9\,$Myr, the cumulative mass of stars escaping from the star system increases by $16\%$. The number of collisions increases only a bit meaning that the most massive object reaches a mass of $4300\rm~ M_\odot$. The black hole formation efficiency $\epsilon_{BH}$ increases by $13\%$; this increase is more due to the escapers than the collisions since the mass of the most massive object increases only by $150\rm~ M_\odot$ while there are several stars that escape from the stellar system.

Model B is one of the densest models. This model is very chaotic, showing a contraction at the beginning. Almost all collisions occur within $1\,$Myr; there are also several escapers during this time, therefore the black hole formation efficiency increases due to both processes. After $1\,$Myr only a few collisions occur, while the number of escapers continues to increase. This means that during the late times the increase of $\epsilon_{BH}$ is dominated more by the number of escapers than by stellar collisions. One possible explanation for why a star cluster is no longer visible in certain observed systems is due to the cluster evaporation as the remaining stars are ejected, this process can be caused by various dynamical interactions within the system, such as binary disruption or interaction with massive objects.



\begin{figure}
    \centering
    \includegraphics[width=\linewidth]{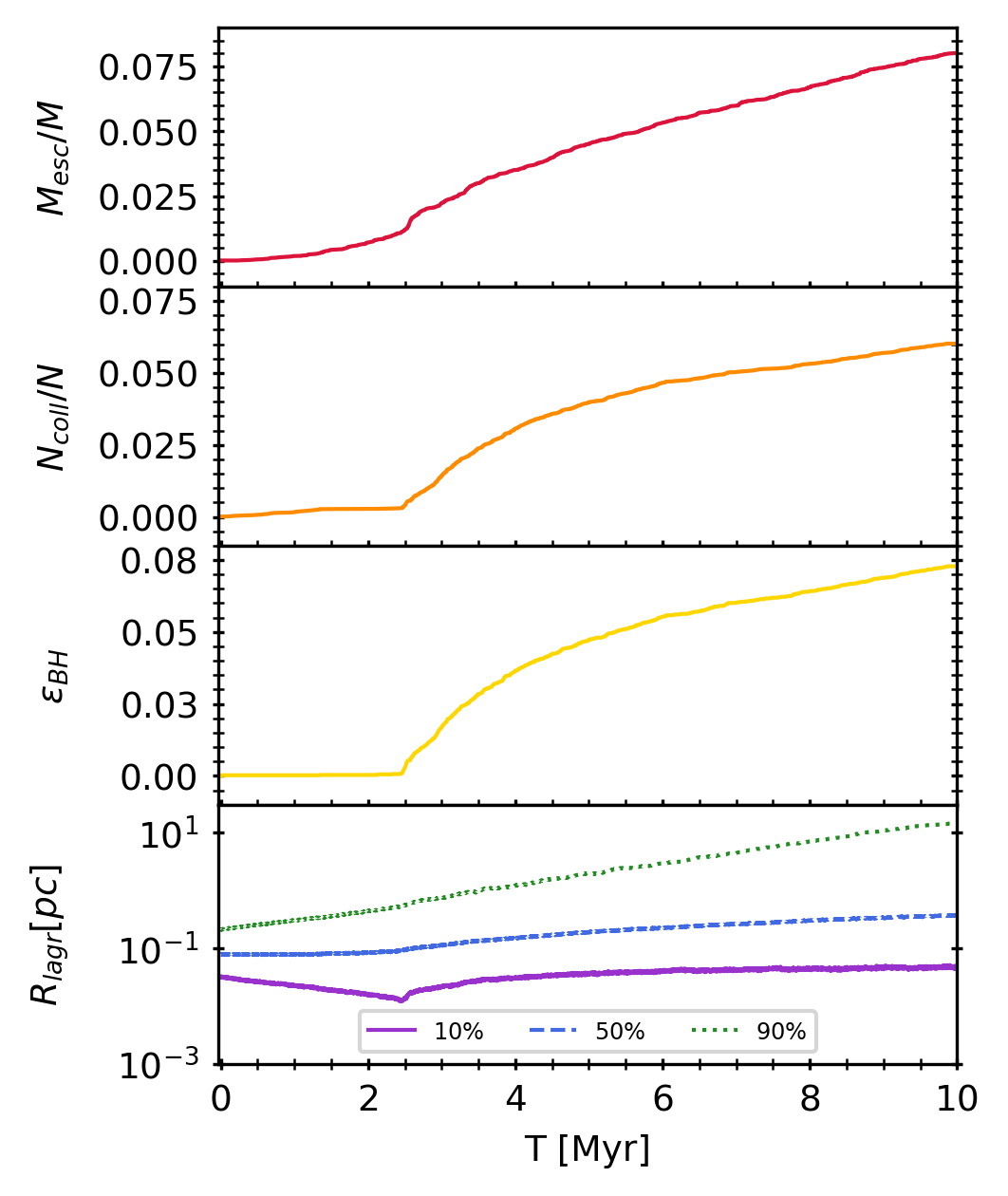} 
    \caption{Evolution of model M for a time period of $10\,$Myr. Panels are the same as in Fig.~\ref{C1Myr}.}
    \label{I10Myr}
\end{figure}

Fig.~\ref{I10Myr} shows the evolution of model M for a time period of $10\,$Myr. The panels are the same as in Fig.~\ref{C1Myr}. The top panel shows that around $8\%$ of the stellar mass is lost due to the escapers. The first middle panel shows around $700$ stellar collisions. The most massive object reaches a mass of $6700\rm~{M_\odot}$. The second middle panel shows that the black hole formation efficiency reaches a value of $7\%$. The Lagrangian radius at $90\%$ shows an expansion over time, the Lagrangian radius at $50\%$ remains almost constant while the Lagrangian radius at $10\%$ shows a decrease until $2.5\,$Myr, followed by an expansion and then remains almost constant. Model M has a long relaxation time ($t_{relax}\approx0.16\,$Myr), so it takes longer for the collapse to occur. The time of core collapse is $t_{cc}\propto 20t_{relax}$, around this time the collisions are triggered and therefore the mass growth of the BH begins, this stellar system shows few collisions before $2.5\,$Myr, forming a massive object of $1670\rm~ M_\odot$. At this time there is a small contraction, so after $2.5\,$Myr the number of collisions increases and several of them occur with the most massive object which increases in mass, this stellar system shows a low black hole formation efficiency after $10\,$Myr.




Stellar gravitational interactions can cause stars to be ejected from the cluster, taking kinetic energy with them and causing the cluster energy to redistribute, leading the cluster to undergo contraction. This collapse of the stellar system is related to the formation of the most massive object since the stars are more likely to collide with each other, therefore there is an increase in the number of collisions. At the same time, many of these collisions occur with a single object. 

Model M is more massive than model B, but also has a larger virial radius, forming a more massive object than model B; nonetheless, its evolution is less chaotic because the collision time scale of model B is shorter than $t_H$, while model M has a collision time scale larger than $t_H$. This can be observed in Fig.~\ref{ic}, since the models are in a different part of the mass-radius parameters space; model B is in a region where the collisions dominate the stellar dynamics, while model M is in a region where stellar collisions are avoided. Also, the chaotic gravitational interaction of model B shows a larger fraction of initial mass loss due to stellar escapers than model M. This means that model B has a higher black hole formation efficiency than model M, while nonetheless, model M forms a more massive object than model B.

\subsection{Black hole formation efficiency}

\begin{figure*}
    \centering
    \includegraphics[width=\textwidth]{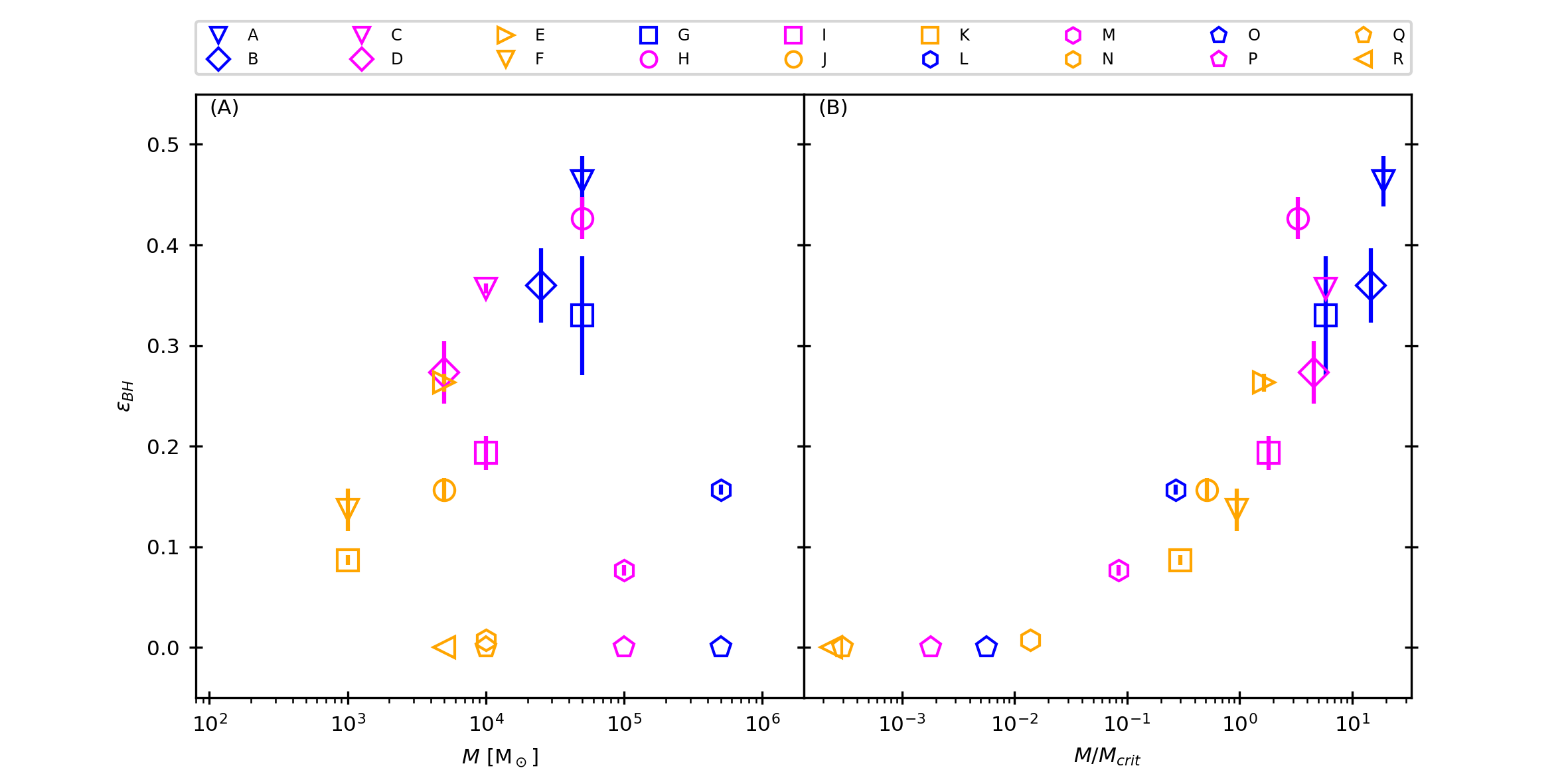}
    \caption{The left panel (A) of Fig.~\ref{eff_wo_Mcrit_10myr} is the black hole formation efficiency computed by $\epsilon_{BH}=(1+M_{NSC}/M_{BH})^{-1}$ (Eq.~\ref{eff_bh}) for $10\,$Myr against the initial mass of the nuclear star cluster $M$. The right panel (B) of Fig.~\ref{eff_wo_Mcrit_10myr} is the black hole formation efficiency computed by Eq.~\ref{eff_bh} against the initial mass of the nuclear star cluster $M$ normalized by the critical mass $M_{crit}$ (Eq.~\ref{eq_mass_crit}) for $t_H=10\,$Myr.}
    \label{eff_wo_Mcrit_10myr}
\end{figure*}

In this subsection, we analyze the black hole formation efficiency ($\epsilon_{BH}$) for the 18 different initial conditions (Table~\ref{ic}), in order to test the scenario for MBH formation proposed by \citet{Escala2021}. Our results are based on the average of three simulations for each of the 18 initial conditions  configurations, with a different random seed to obtain reliable statistics and error estimates.

On the left panel (A) of Fig.~\ref{eff_wo_Mcrit_10myr}, we display the black hole formation efficiency as a function of the initial mass of the NSCs. At first glance, there seem to be three trends; according to the colors, they are ordered as orange, magenta, and blue from left to right. However, in order to test the global instability proposed by \citet{Escala2021}, it  is useful to normalize by the critical mass ($M_{crit}$); we use this mass to normalize the mass of the clusters since  we need to quantify how near or far away they are from the collision line in Fig.~\ref{ic}.

On the right panel (B) of Fig.~\ref{eff_wo_Mcrit_10myr}, we display the black hole formation efficiency against the initial mass of the nuclear cluster normalized by the critical mass ($M_{crit}$) at $10\,$Myr. The normalization by critical mass is helpful to show a clear trend in the efficiency of black hole formation, where the models with the highest efficiency are also the models where the collision time scale is shorter than $t_H$ and the models with lower black hole formation efficiency are those which have a collision time scale larger than $t_H$. Model A shows the highest efficiency of around $46\%$. Model B has $\epsilon_{BH}\approx36\%$. Models C and D show a black hole formation efficiency of around $35\%$ and $27\%$, respectively.  Models E and F have $\epsilon_{BH}\approx27\%$ and $\epsilon_{BH}\approx15\%$, respectively. Model G shows an efficiency of around $33\%$. Model H has $\epsilon_{BH}\approx44\%$, and Model I $\epsilon_{BH}\approx19\%$. Models J and K show a black hole formation efficiency of around $17\%$ and $9\%$, respectively, while models L and M show $\epsilon_{BH}\approx16\%$, and $\epsilon_{BH}\approx8\%$, respectively. The black hole formation efficiency is less than $1\%$ for models O and N. $\epsilon_{BH}=0\%$ for models P, Q, and R. Our models correctly quantify the expected behavior of the new proposed scenario of \citet{Escala2021} of MBH formation through runaway collisions in NSCs, where models with $t_{coll}<t_{H}$, show a global instability where a great percentage of the stellar  mass collapses into a single object, while $t_{H}<t_{coll}$ shows stellar systems that practically avoid collisions.

According to Fig.~\ref{ic} we described different regions of the mass-radius parameter space (Fig.~\ref{ic}). Models in the region where the collision time is longer than the simulation time ($t_H$) show the lowest black hole formation efficiencies e.g. models O, P, Q, and R with $\eqsim 0\%$, including model N with $\sim 1\%$, so in this region, collisions are almost entirely avoided. Models K and M show a slightly higher, but still low black hole formation efficiency, reaching values less than $10\%$. They still are in the parameter space where $t_{coll}>t_{H}$ but show a bit more collisions since the ratio between the timescales is reduced. They are followed by models F, I, J, and L which show a black hole formation efficiency of less than $20\%$. They are near the lines described by the collision Eq.~\ref{coll_ts}. When the collision time is shorter than the simulation time ($t_H$) the clusters show the highest black hole formation efficiency where collisions dominate the stellar dynamics of the system, e.g models D and E with a black hole formation efficiency larger than $20\%$; also models B, C, and G show a black hole formation efficiency above $30\%$ and 
models A and H have a black hole formation efficiency of around $50\%$. Note that the error bar for the black hole formation efficiency for models N and O is quite small (practically zero) and models P, Q, and R do not show an error bar due to these systems not having collisions.

We summarize our results in Table~\ref{table2}. We provide in the first column the model ID, in the second column the cumulative mass of the escapers ($M_{esc}$), in the third column the final black hole mass ($M_{BH}$), in the fourth column the final mass of the nuclear star cluster ($M_{NSC}$), in the fifth column the final mass of the central massive object ($M_{CMO}$) and the black hole formation efficiency ($\epsilon_{BH}$) is in the last column.

\begin{table*}
    \centering
    \caption{$M_{esc}$ is the cumulative mass of the escapers, the final black hole mass is $M_{BH}$, $M_{NSC}$ is the final mass of the nuclear star cluster, $M_{CMO}$ is the sum of $M_{BH}$ and $M_{NSC}$, and the black hole formation efficiency is $\epsilon_{BH}=(1+M_{NSC}/M_{BH})^{-1}$. All quantities are measured for a time period of $10\,$Myr. For the cases marked with $^*$, an error estimate was not possible, as all conducted simulations produced the same result. While the uncertainty in these cases  should be low, a precise quantification was not feasible.}
    \begin{tabular}{cllllc}\hline\hline
      Models ID & $M_{esc}~\rm[{M_\odot}]$&$M_{BH}~\rm[{M_\odot}]$  & $M_{NSC}~\rm[{M_\odot}]$  & $M_{CMO}~\rm[{M_\odot}]$ & $\epsilon_{BH}$ \\\hline
A & $28533\pm1477$ & $9917\pm143$ & $11550\pm1353$ & $21467\pm1477$ & $46.33\pm2.5\%$\\  
B & $14300\pm687$ & $3850\pm324$ & $6850\pm795$ & $10700\pm687$ & $36.00\pm3.7\%$\\  
C & $5370\pm42$ & $1660\pm14$ & $2970\pm57$ & $4630\pm42$ & $35.67\pm0.5\%$\\ 
D & $2767\pm111$ & $610\pm43$ & $1623\pm153$ & $2233\pm111$ & $27.33\pm3.1\%$\\
E & $2061\pm56$ & $778\pm20$ & $2161\pm71$ & $2939\pm56$ & $26.33\pm0.9\%$\\  
F & $372\pm5$ & $85\pm12$ & $542\pm16$ & $628\pm5$ & $13.67\pm2.1\%$\\  
G & $26183\pm1569$ & $7800\pm788$ & $16017\pm2357$ & $23817\pm1569$ & $33.00\pm5.9\%$\\  
H & $23137\pm1966$ & $11397\pm143$ & $15380\pm1722$ & $26863\pm1966$ & $42.67\pm2.1\%$\\  
I & $4253\pm183$ & $1110\pm83$ & $4637\pm252$ & $5747\pm183$ & $19.33\pm1.7\%$\\  
J & $1459\pm22$ & $545\pm42$ & $2996\pm64$ & $3541\pm22$ & $15.67\pm1.2\%$\\  
K & $327\pm8$ & $59\pm2$ & $614\pm10$ & $673\pm8$ & $8.67\pm0.5\%$\\ 
L & $75517\pm1812$ & $67450\pm2585$ & $357433\pm2121$ & $424883\pm1812$ & $15.67\pm0.5\%$\\  
M & $7850\pm283$ & $6987\pm246$ & $85163\pm130$ & $92150\pm283$ & $7.67\pm0.5\%$\\  
N & $234\pm44$ & $71\pm5$ & $9695\pm49$ & $9766\pm44$ & $0.73\pm0.0\%$\\  
O & $450\pm178$ & $100\pm0^*$ & $499450\pm178$ & $499550\pm178$ & $0.02\pm0.0\%$\\  
P &$27\pm17$&$0\pm0^*$&$99973\pm17$&$99973\pm17$&$0.00\pm0.0\%^*$\\
Q & $0\pm0^*$ & $0\pm0^*$ & $10000\pm0^*$ & $10000\pm0^*$ & $0.00\pm0.0\%^*$ \\
R & $0\pm0^*$ & $0\pm0^*$ & $5000\pm0^*$ & $5000\pm0^*$ & $0.00\pm0.0\%^*$ \\
    \hline   
    \end{tabular}
    \label{table2}
\end{table*}

Fig.~\ref{eff_wo_Mcrit_10myr} (B) shows that our models quantify correctly the expected behavior of the proposed new scenario of MBH formation through runaway collisions in NSCs. We recall that we considered simplified and more compact toy models in our simulations to make the calculations computationally feasible, considering different regions of the parameter space that includes both regions where collisions would be expected to be very efficient and very inefficient (see Fig. 1). While of course, it has to be checked in further detail, one may expect that a similar transition might occur for any model that fulfills the condition $t_H>t_{coll}$ that we explore here. 

\section{Observational counterpart}

In this section, we perform calculations to determine the collision timescale, critical mass, and black hole formation efficiencies for some observed galactic nuclei. We examined several properties of the galactic centers, including the black hole mass, the nuclear stellar mass, and the effective radius. Since MBHs and NSCs are found together in galactic nuclei, it is likely that they share a common formation process \citep{Georgiev2016, Neumayer2020, Escala2021}. For simplicity, we here assume that the initial mass of the galactic center was the sum of the masses of the MBH and NSC. We also assume that the NSCs initial effective radius was ten times smaller than its current size \citep{Banerjee2017}. Besides we assume that the stellar system consists of $1~\rm{M_\odot}$ and $1~\rm{R_\odot}$ stars. Finally, the value of $t_H$ for the galactic centers must be of the order of $\sim 10^9\,$yr considering their typical formation times \citep{Walcher2005, Rossa2006}.

With these parameters, we are able to compute the critical mass, the black hole formation efficiency, as well as the collision timescale. Our estimated values depend on several factors, which can vary depending on the adopted assumptions and approximations. Therefore, it is important to keep in mind that our simplified estimates may not precisely match the results of more detailed studies. We summarize the principal properties of the galactic centers and our calculations in table~\ref{table3}.

\begin{table*}
    \centering
    \caption{(1) Galaxy name. Observational properties of the galactic center, columns: (2, 3) BH mass and reference, (4, 5) NSC mass and reference, (6, 7) NSC effective radius and reference. Estimation of galactic centers properties, columns: (8) collision timescale, (9) critical mass (Eq.~\ref{eq_mass_crit}), (10) black hole formation efficiency (Eq.~\ref{eff_bh}). The symbol $<$ denotes upper mass limits.\\ 
     References: (1) \citet{Do2019}, (2) \citet{Schodel2014}, (3) \citet{Feldmeier2014}, (4) \citet{Nguyen2018}, (5) \citet{Nguyen2019}, (6) \citet{Neumayer2012}, (7) \citet{Georgiev2016}, (8) \citet{Walcher2005}, (9) \citet{Sarzi2001}, (10) \citet{Pechetti2020}, (11) \citet{Emsellem1999}, (12) \citet{Graham2009}, (13) \citet{Nguyen2022}, (14) \citet{Barth2009}, (15) \citet{denBrok2015}, (16) \citet{Peterson2005}, (17) \citet{Thater2017}, (18) \citet{Gebhardt2011}, (19) \citet{Gnedin2014}, (20) \citet{Graham2008}, (21) \citet{Fusco2022}.}
    \begin{tabular}{lrlrlclcrrr}\hline\hline
      Galaxy &  $M_{BH}~\rm[{M_\odot}]$& Ref. & $M_{NSC}~\rm[{M_\odot}]$& Ref. & $R_{e,NSC}~\rm[{pc}]$& Ref.& \qquad &$t_{coll}~\rm[{Myr}]$& $M_{crit}~\rm[{M_\odot}]$  &$\epsilon_{BH}$ \\
      \hspace{.3cm}(1) & (2)\hspace{.5cm} & (3) & (4)\hspace{.5cm} & (5) & (6) & (7) &  & (8)\hspace{.5cm} & (9)\hspace{0.45cm} & (10)\hspace{.05cm} \\\hline
    Milky Way & $4.0\times 10^{6}$ &(1)& $3.0\times 10^{7}$ &(2)& $4.2$ &(2, 3)&   & $828$  & $3.0\times10^7$  & $11.76\%$  \\
    NGC 205 & $ 6.8\times10^3$ &(4)& $1.0\times10^6$ &(5)& $1.3$ &(5)& & $883$ & $9.3\times10^5$ & $0.67\%$ \\
    NGC 221& $ 2.5\times10^6$ &(5)& $1.7\times10^7$ &(5)& $4.4$ &(5)&  & $1962$ & $3.0\times10^7$ & $13.16\%$ \\
    NGC 428 & $<7.0\times10^4$ &(6)& $2.6\times10^6$ &(7, 8)& $1.2$ &(7)& & $300$ & $1.2\times10^6$ & $2.62\%$ \\
    NGC 1042 & $<3.\times10^6$ &(6)& $3.2\times10^6$ &(8)& $1.3$ &(7)&  & $152$ & $1.8\times10^6$ & $48.39\%$ \\
    NGC 1493 & $<8.0\times 10^{5}$ &(6)& $3.5\times10^{6}$ &(7)& $3.6$ &(7)&   & $4745$  & $1.2\times10^7$  & $18.43\%$\\
    NGC 2139 & $<4.0\times10^{5}$ &(6)& $2.7\times10^{7}$ &(7)& $14.8$\hspace{.1cm} &(7)&   & $53700$  & $4.0\times10^8$  & $1.44\%$\\
    NGC 2787 & $  4.1\times10^7$ &(20)& $<7.0\times10^7$ &(9)& $5.1$ &(10)&  & $324$ & $5.2\times10^7$ & $36.94\%$ \\
    NGC 3115 & $9.1\times10^{8}$ &(11)& $7.2\times10^{6}$ &(12)& $6.6$ &(10)& & $36$  & $1.0\times10^8$  & $99.22\%$\\ 
    NGC 3423 & $<7.0\times10^{5}$ &(6)& $1.9\times10^{6}$ &(7)& $2.2$ &(7)&  & $1779$  & $3.9\times10^6$  & $26.52\%$\\
    NGC 3593 & $2.4\times10^6$ &(13)& $1.7\times10^7$ &(13)& $5.1$ &(10, 13)& & $2916$ & $3.9\times10^7$ & $12.57\%$\\
    NGC 3621 & $<3.0\times10^{6}$ &(14)& $6.5\times10^{6}$ &(7)& $1.8$ &(7)&  & $257$  & $3.9\times10^6$  & $31.35\%$  \\
    NGC 4395 & $ 3.6\times10^5$ &(15, 16)& $2.0\times10^6$ &(7, 15)& $1.5$ &(7)& & $65$ & $1.8\times10^6$ & $15.25\%$ \\
    NGC 4414 & $  <1.5\times10^6$ &(17)& $1.2\times10^8$ &(7)& $26.5$\hspace{.1cm} &(7)&  & $64450$ & $2.0\times10^9$ & $1.25\%$ \\
    NGC 4486 & $6.6 \times 10^{9}$ &(18)& $<2.0\times10^{8}$ &(6)& $7.4$ &(19)&  & $3$  & $1.3\times10^8$  & $97.06\%$\\
    NGC 5055 & $8.5\times10^8$ &(20)& $1.7\times10^6$ &(10)& $13.5$\hspace{.1cm} &(10)&  & $482$ & $5.2\times10^8$ & $99.81\%$\\
    NGC 5102 & $9.1 \times 10^{5}$ &(5)& $7.3\times10^{7}$ &(4)& $26.3$\hspace{.1cm} &(4)&  & $112944$  & $1.7\times10^9$  & $1.23\%$\\
    NGC 5206 & $6.3\times10^{5}$ &(5)& $1.5\times10^{7}$ &(4)& $8.1$ &(4)&   & $15294$  & $1.0\times10^8$  & $3.94\%$ \\
    NGC 7424 & $ < 4.0\times10^5$ &(6)& $1.0\times10^6$ &(6, 7)& $6.8$ &(7)&  & $57860$ & $2.1\times10^7$ & $28.57\%$ \\
    NGC 7713 & $7.5\times10^6$ &(21)& $4.0\times10^5$ &(10)& $1.1$ &(10)&  & $74$ & $1.4\times10^6$ & $94.89\%$ \\ 
    \hline   
    \end{tabular}
    \label{table3}
\end{table*}

\begin{figure}
    \centering
    \includegraphics[width=\linewidth]{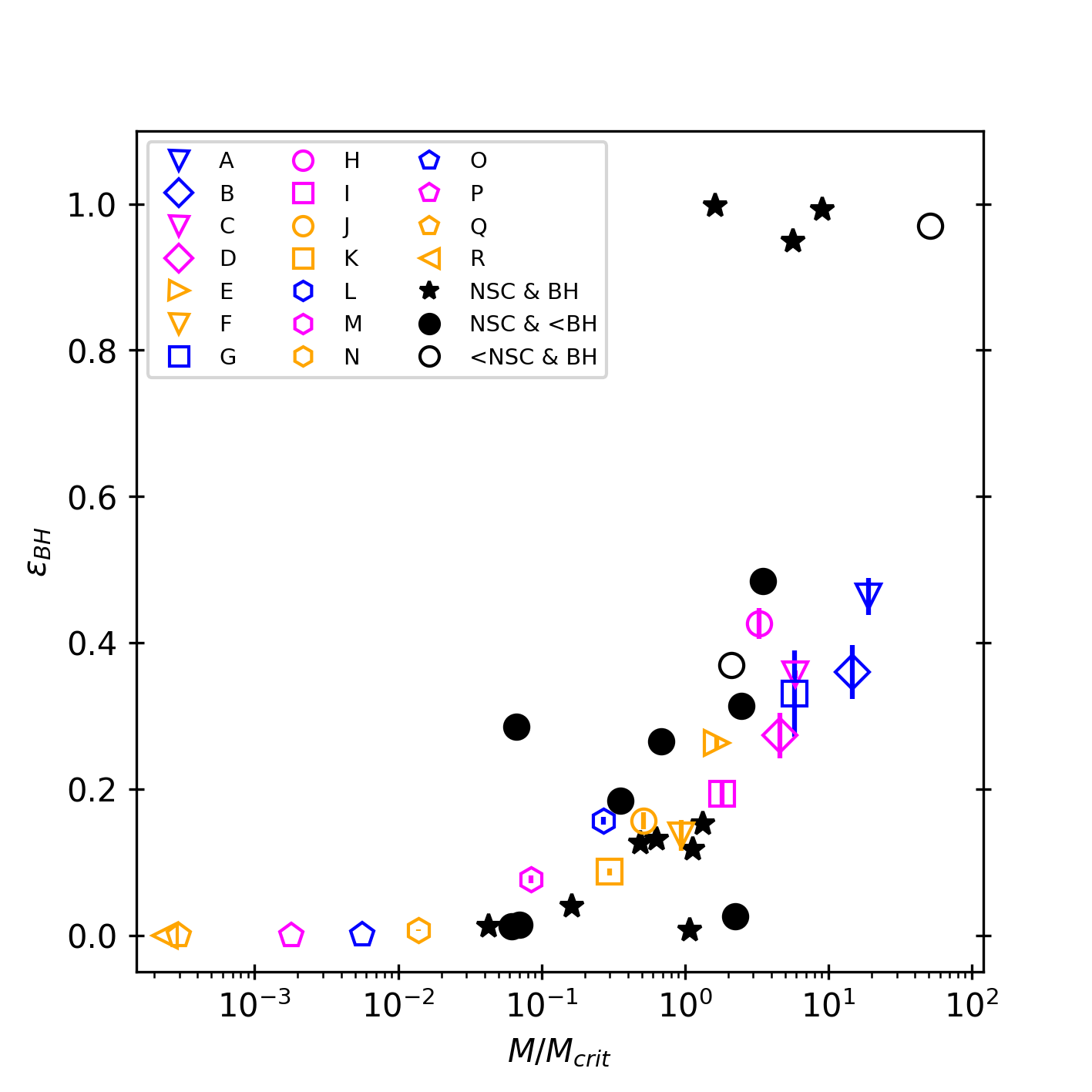} 
    \caption{The black hole formation efficiency computed by Eq.~\ref{eff_bh} against the initial mass of the nuclear star cluster $M$ normalized by the critical mass $M_{crit}$ (Eq.~\ref{eq_mass_crit}) for $t_H=10\,$Myr in our models and the assumed initial mass normalized by the critical mass for $t_H=10^9\,$yr for the galactic centers. Objects that have both NSCs and BHs masses are denoted by black stars, while those with upper limits on BHs masses are denoted by black circles, and those with upper limits on NSCs masses are denoted by white circles.}
    \label{BH_eff_2}
\end{figure}

In Fig.~\ref{BH_eff_2} we show the black hole efficiency as a function of the ratio of the total mass normalized by the critical mass of the galactic center for $t_H=10^9\,$yr, along with the data from the simulations for comparison. The observational values are depnd on multiple factors that may differ based on the assumptions and  approximations applied to the observations. Besides, we recall that we made several assumptions calculating the properties of the observed galactic nuclei. However, after conducting an analysis and comparing it with our simulations, we find a significant level of agreement, thereby supporting our proposed scenario for the formation of black holes through collisions in nuclear stellar clusters.

The efficiency of black hole formation varies depending on the relative masses of the NSC and the MBH at the galactic center. When the NSC is more massive than the MBH in galaxies like NGC 221, NGC 1493, NGC 2139, NGC 3593, NGC 4414, NGC 5102, and NGC 5206, it implies that $t_H<t_{coll}$ and leads to a black hole formation efficiency of less than $20\%$. This trend is consistent with our simulations, including models F, I, J, K, L, M, and N. NGC~3423 is an exception, with $t_H<t_{coll}$ and a black hole formation efficiency higher than $20\%$, but it still aligns with the simulation trend. The Milky Way and NGC 4395 have $t_{coll}<t_H$, and their black hole formation efficiencies are lower than $20\%$ but higher than $10\%$. However, they still agree with the trend of the simulations. 

In contrast, when the MBH dominates the galactic nuclei, such as in NGC 3115, NGC 4486, NGC 5055, and NGC 7713, we have $t_{coll}<t_H$. In these galactic nuclei, a jump in the black hole formation efficiency occurs, probably due to an evolutionary process like evaporation. Likewise, we show in Fig.~\ref{C10Myr} that the rise of the black hole formation efficiency at late times in the simulation is mainly influenced by the number of escapers. Thus the disappearance of visible nuclear stellar clusters in some observed systems is possibly due to the ejection of the remaining stars causing the cluster to evaporate.

Galactic centers with comparable masses between the NSC and MBH, such as NGC 1042, NGC 2787, and NGC 3621, which have $t_{coll}<t_H$, result in black hole formation efficiencies of approximately $\approx30$-$50\%$. These galactic centers align well with our models A, B, C, G, and H.

Despite having galactic centers with $t_{coll} < t_H$, the black hole formation efficiencies of NGC 205 and NGC 428 are less than $3\%$, due to the low mass of their black holes ($\sim 10^3-10^4~\rm{M_\odot}$) compared to the NSC mass ($\sim10^6~\rm{M_\odot}$). This possibly indicates that their current effective radius is more similar to the radius at formation, deviating from the assumed factor of 10. This is a possible uncertainty inherent in our assumptions. On the other hand, NGC 7424 has a black hole formation efficiency of approximately $30\%$, despite having $t_H < t_{coll}$. This value is similar to a case like NGC 3423 but does not align with the trend observed in our simulations, possibly because this system has only an upper limit on the black hole mass. Therefore these galactic centers may deviate from the expected trend.

The observed properties of the galactic nuclei are subject to various factors that may differ depending on the methods used during the observations. Additionally, we made several assumptions calculating the properties of the observed galactic centers. Despite this, our analysis comparing observations to simulations shows a significant level of agreement, thus supporting the proposed scenario by \citet{Escala2021} for the formation of black holes through collisions in nuclear stellar clusters.

\section{Conclusions}

In this work, we investigated the behavior of nuclear stellar clusters and their ability to form massive black holes under runaway collisions, testing the global collapse scenario presented in \citet{Escala2021}, motivated by the observation of NSCs in the regime where collisions are not dynamically relevant in the global  evolution of the systems, and the observation of MBHs in the regime where collisions are dynamically relevant. Thus \citet{Escala2021} proposed that stellar configurations with $t_{coll}<t_H$ must show a global instability of the stellar system which collapses a great part of the mass into one single object, while a stellar system with $t_{coll}>t_H$ must avoid runaways collisions. To explore this new scenario regardless of the computational limitations, since it is quite expensive to run simulations with a larger number of stars ($N<10^6$), we here present models of compact NSCs (i.e with low critical mass, thus feasible to explore with N-body simulations) with short simulation times that fulfill the condition $t_{coll}<t_H$ (i.e stellar system above the collisions line described by Eq.~\ref{coll_ts} in Fig.~\ref{ic_10myr}) as a proof of concept that the expected type of instability occurs at least for simplified systems. Our results support the further exploration of this instability in the future and also motivate the possible relevance of this type of instability in real NSCs, providing a potential formation channel for very massive objects.

We analyzed the Lagrangian radii, the mass loss due to stars escaping from the system, the number of stellar collisions and the black hole formation efficiency. We performed an analysis of 18 different models covering different regions of the mass-radius parameter space, described in Fig.~\ref{ic}. If $t_{coll}<t_{H}$, the stellar system becomes unstable under collisions, causing a chaotic collapse, ejecting several stars, and forming a massive object. On the other hand, if $t_{coll}>t_{H}$, the system practically avoids almost all of the collisions and experiences very few escapes. When the system is closer to the curve defined by Eq.~\ref{relax_ts}, fewer collisions will occur, while if the model is closer to the curve from Eq.~\ref{coll_ts}, more collisions will happen. The goal of this work was to provide a proof of concept of the critical transition derived by \citet{Escala2021}; for computational reasons, we could not model real nuclear star clusters with masses of the order $10^7$~M$_\odot$, but we considered smaller and more compact systems following shorter evolutionary times, leading overall to a reduction of the critical mass scale to $10^3-10^5\rm~ M_\odot$ which made the numerical modeling feasible.

We define the efficiency $\epsilon_{BH}$ as the ratio of black hole mass to final stellar mass. The models with more chaotic evolution show that at the beginning the black hole formation efficiency is dominated by stellar collisions, while at late times, the increase of the black hole formation efficiency is quickly dominated by the mass loss due to stellar escapes. 

Systems in the region where collisions are avoided have a longer relaxation time; these systems expand rapidly before there is time for collisions to occur, showing a very low black hole formation efficiency that in some cases can be equal to zero. On the other hand, the systems in the region where the collisions are relevant do not have enough time to expand before the collisions dominate the stellar dynamics, forming a very massive object in the center and reaching high black hole formation efficiencies. The extreme models A and H have the same initial mass of $5\times10^4\rm~ M_\odot$, with a virial radius of $0.005\rm~ pc$ and $0.01\rm~ pc$, respectively; these systems form a massive object of $M_{BH}\approx10^4\rm~ M_\odot$ and lose around of $50-60\%$ of the initial mass due to the escapers. The black hole formation efficiency of these systems is $\epsilon_{BH}\approx50\%$, which is the highest of all our models. On the other hand, models where collisions are avoided, show a low black hole formation efficiency; particularly models P, Q, and R have a long relaxation time so they take longer to core collapse and until the end of our simulations they show no collisions (i.e. the black hole formation efficiency is $0\%$). There are systems that do not cross the line described by Eq.~\ref{coll_ts} but are close to it (e.g models B, C, D, E, and G). These show a black hole formation efficiency higher than $20-30\%$ while other systems (e.g F, I, J, K, L, and M) further away from the collision line show lower black hole formation efficiencies of around $10-20\%$. We find, that the black hole formation efficiency is high in the parameter space where the collisions are relevant for the global instability, and the violent behavior of the NSC before the cluster expansion allows the formation of a MBH, besides that in our simulations $\rm t_{relax} < t_{coll}$, leading to an even simpler scenario and more similar to the one originally proposed in \citet{Escala2020}. Here we explored setups where $t_{H}=10\,$Myr, with the ideal conditions to be only proof of concept of global collapse triggered by collisions. 

So the occurrence of a transition in the black hole formation efficiency was clearly demonstrated within our toy models. As mentioned above, in real NSCs the critical mass scale will be larger of the order $10^7$~M$_\odot$, and they also have larger masses of up to a few $10^8\rm~ M_\odot$. Assuming that a black hole formation efficiency of $50\%$ is possible, such systems could potentially form supermassive black holes with up to $\sim 10^8\rm~ M_\odot$. The Universe has an age of $13.6\,$Gyr, which means that even more extended systems with longer collision and relaxation time scales can go through this global collapse. We further conducted a preliminary analysis by comparing our simulations with observed NSCs. The results indicate a significant level of agreement, supporting the viability of our proposed scenario of black hole formation via collisions in nuclear stellar clusters. However, further investigation and in-depth analysis are required to fully understand the implications of this scenario.

Some studies suggest that massive seeds ($\sim 10^5\rm~M_\odot$) are needed to explain the observed supermassive black holes at high redshift \citep{Pezzulli2016, Valiante2016, Sassano2021, Trinca2022}. Our chaotic models reach black hole masses of the order of $10^3-10^4\rm~ M_\odot$, results consistent with the simulations of \citet{Devecchi2009, Sakurai2017, Reinoso2018}. In realistic more massive systems in principle the formation of even more massive central objects is thus expected. In general, we expect that for dense models where the parameter space is dominated by collisions higher black hole masses are reached when the long available times are taken into account ($>10^5 \rm~ M_\odot$)
\citep{Lee1987, Quinlan1990, Davies2011, Stone2017}. If NSCs are born with a radius such that their initial mass is $\rm 10 \, M_{crit}$, they can lead to the formation of massive objects of $\sim 10^5\rm~M_\odot$ assuming a $50\%$ efficiency. 

The scenario proposed here can also be applied to globular clusters. They typically fall into the mass range of $10^5$-$10^6\rm~{M_\odot}$ \citep{Tremou2018}, have radii of a few parsecs and an age of $10^{10}\,$yr, leading to a critical mass of around $10^7\rm~{M_\odot}$. Therefore, the mass of these systems is $10^{-2}$-$10^{-1}$ times the critical mass at formation, i.e, the initial mass of the cluster is $\sim 1$-$2$ orders of magnitude below the collision line define by Eq.~\ref{coll_ts}. Considering Fig.~1 of \citet{Escala2021}, these stellar systems are expected to show a low black hole formation efficiency. It has been suggested that the mass of an intermediate-mass black hole is between $10^2$-$10^4\rm~{M_\odot}$  \citep{Noyola2006, Feng2011}, resulting in a low $e_{BH}\sim 0.1$-$1\%$, which would be compatible with this scenario.

Eventually observations of NSCs with the James Webb Space Telescope (JWST)\footnote{JWST: \url{ https://webb.nasa.gov}} will be possible at high redshift ($3\leq z\leq 8$) \citep{Renzini2017}, and similar for the Extremely Large Telescope (ELT)\footnote{ELT: \url{ https://elt.eso.org}} which will be equipped with MICADO, the Multi-AO Imaging Camera for Deep Observations at near-infrared wavelengths. The high spatial resolution of MICADO will allow the spheres of influence to be resolved with greater precision, considerably increasing the available surveys of supermassive black holes masses covering a black hole mass range of $10^5$-$10^7\rm~ M_\odot$ \citep{Davies2018}. The high spatial resolution of MICADO will allow resolving the sphere of influence at a 5 times larger distance than the current instruments, also 2 times larger than the JWST. MICADO will observe many additional NSCs and determine many supermassive black hole masses that will allow the efficiency of black hole formation to be determined with high precision. The large observational data set could be compared with the results of our models. Our models suggest that NSCs could form a more massive object than $10^5\rm~ M_\odot$ through runaway collisions. It is difficult to make direct observations of the formation process of MBH, especially in this stellar compact configuration, gravitational waves are also expected to occur at the galactic center \citep{Rees1984}, thus the detection of gravitational waves using the interferometers such as  LIGO, Virgo, and Kagra or in the future LISA, and ET would be fundamental \citep{Fragione2020, Fragione2022}. These observations will also help to probe the new formation scenario proposed here, by providing an accurate estimation of black hole masses for a large range of different clusters.


\subsection{Potential caveats for future improvement}


As mentioned, NSCs live in the centers of galaxies and it has been suggested that the formation of NSCs is due to the accretion of globular clusters, which fall to the center by dynamic friction \citep{Antonini2012}. This mechanism is called a cluster-inspiral and is generally invoked as an explanation for the rotation observed in NSCs \citep{Seth2008}. Therefore, including rotation in NSC simulations will be important when developing more realistic simulations, since the presence of rotation in the spherical models leads to a deformation in the outer zone of the cluster, appearing in a non-spherical distribution \citep{Varri2012, Lupton1987}. Rotation in stellar systems slightly reduces collisions due to the ordered motion, however, rotation also causes a flattening of the cluster, increasing the density and number of collisions \citep{Vergara2021}.

Another important simplification in this work is to simulate clusters only with equal-mass stars. Stellar populations are complex since they are born with an initial distribution of the masses of their stars that is called the initial mass function (IMF) \citep{Salpeter1955}. The IMF is similar within the Milky Way and nearby star-forming regions \citep{Kroupa2001, Chabrier2003}. The IMF implies mass segregation, as massive objects tend to fall toward the center while light objects move outward \citep{Baumgardt2008}, which explains the depletion of low-mass stars \citep{Aarseth1972}, However, we expect a violent evolution, which collapses a great part of the initial mass into a single object, regardless of stellar mass. The evolution of clusters depends strongly on their primordial binaries since a small fraction of binary systems can play a crucial role in the dynamics of the clusters \citep{Goodman1989, Portegies2001}.

Our simulations do not take into account stellar evolution, so our NCSs are dominated by gravity. However, it is important to consider that due to the mass loss produced by stellar winds and supernova explosions, these winds lead to a strong expansion at the beginning of the cluster evolutionary process (after about $10\,$Myr) \citep{Applegate1986, Chernoff1987, Chernoff1990, Fukushige1995}. Due to mass segregation, the most massive stars or stellar black holes sink into the cluster center, where the stars often form hard binaries with high eccentricities, implying that this dense stellar configuration is a source of gravitational waves. Therefore it is necessary to use post-Newtonian N-body dynamics to successfully solve this scenario, as done for example in the works of \citet{Blanchet2006, Brem2013, Rizzuto2021, Rizzuto2022, Arca-Sedda2021}. Another simplification in this work is the assumption that after the collisions the new star reaches hydrostatic and thermal equilibrium quickly. However, due to the high-speed encounters of the stars at the core of the NSC, the time scale of successive collisions must be shorter than the thermal timescale of the new stars. This could then lead to problems in the formation of a very massive object, for example due to increased mass loss in a non-thermalized system. The potential relevance of this problem, depending on the frequency of the collisions, thus should be investigated in further detail \citep{Freitag2006}.

Numerical simulations of a large number of particles require a lot of time and computational resources. However, a few such very large simulations have been performed with the DRAGON simulations of globular clusters that include $10^6$ stars \citep{Wang2016} using \href{https://github.com/nbodyx/Nbody6ppGPU}{{\sc nbody6++gpu}} \citep{Wang2015}. While being computationally very expensive, such simulations will be important to further test the proposed new scenario based on runaway collisions in NSCs as a mechanism to form MBHs \citep{Escala2021} for a greater number of stars ($N=10^5-10^6$) and more massive clusters ($\sim 10^6-10^8\rm~ M_\odot$).

Including gas in NSCs is another interesting option, especially at higher redshifts, where gas in galaxies can account for up to 80\% of the baryonic mass for some extreme cases \citep{Molina2019}. The dissipative nature of gas should enhance stellar collision, for example through dynamical friction \citep{Ostriker1999, Escala2004} but the presence of gas in a cluster could also delay the formation of a MBH due to stellar collisions since stars reach higher velocities, so it is more difficult for close encounters to occur. These systems also have a longer relaxation time; however, if the simulation is long enough these systems could form a MBH, since the gas limits the expansion of the cluster, allowing more stars to remain in the cluster, thus forming a more massive BH than a gasless system \citep{Reinoso2020}. Also, the high density in systems with gas must allow the formation of MBHs with masses of the order $\lesssim 10^5\rm~ M_\odot$ \citep{Davies2011}. Quite similarly, the interaction between the gas and the protostars can also lead to the formation of massive objects \citep{Boekholt2018, Regan2020, Chon2020, Schleicher2022, Schleicher2023, Reinoso2023}.



\section*{Acknowledgements}
We thank the anonymous referee for a very constructive report that helped to improve our manuscript. MCV acknowgledge funding through ANID (Doctorado acuerdo bilateral DAAD/62210038) and DAAD (funding program number 57600326). MCV, DRGS and AE acknowledge financial support from Millenium Nucleus NCN19$\_$058 (TITANs) and also support from the Center for Astrophysics and Associated Technologies CATA (FB210003). AE also acknowledge financial support from FONDECYT Regular grant $\#$1181663. DRGS also acknowledge financial support from FONDECYT Regular grant $\#$1201280 and through the Alexander von Humboldt - Foundation, Bonn, Germany. BR acknowgledges funding through ANID (CONICYT-PFCHA/Doctorado acuerdo bilateral DAAD/62180013) and DAAD (funding program number 57451854). These resources made the presented work possible, by supporting its development. 


\section*{DATA Availability}
The data underlying this article will be shared on reasonable request to the corresponding author.








\appendix

\section{Computational accuracy}\label{App1}

One of the key challenges in simulating N-body systems is maintaining the accuracy of the computations over long periods of time. One way to track the accuracy of the computations is to check the conservation of energy errors. The energy conservation error in N-body simulations can vary widely depending on the specific simulation being performed and the parameters used. In general, acceptable levels of energy conservation error are in the order of $1\%$ or less. We ran the simulations with \href{https://github.com/nbodyx/Nbody6ppGPU}{{\sc nbody6++gpu}} \citep{Wang2015}, using the following parameters, time step factors {\sc $\rm ETAR=0.01$} and {\sc $\rm ETAI=0.01$}, optimal neighbor number {\sc $\rm NNBOPT$} usually vary between 50-200, the criterion for regularization search {\sc $\rm DTMIN=1.e-4$} and {\sc $\rm RMIN=1.e-4$}.

\begin{figure}
    \centering
    \includegraphics[width=\linewidth]{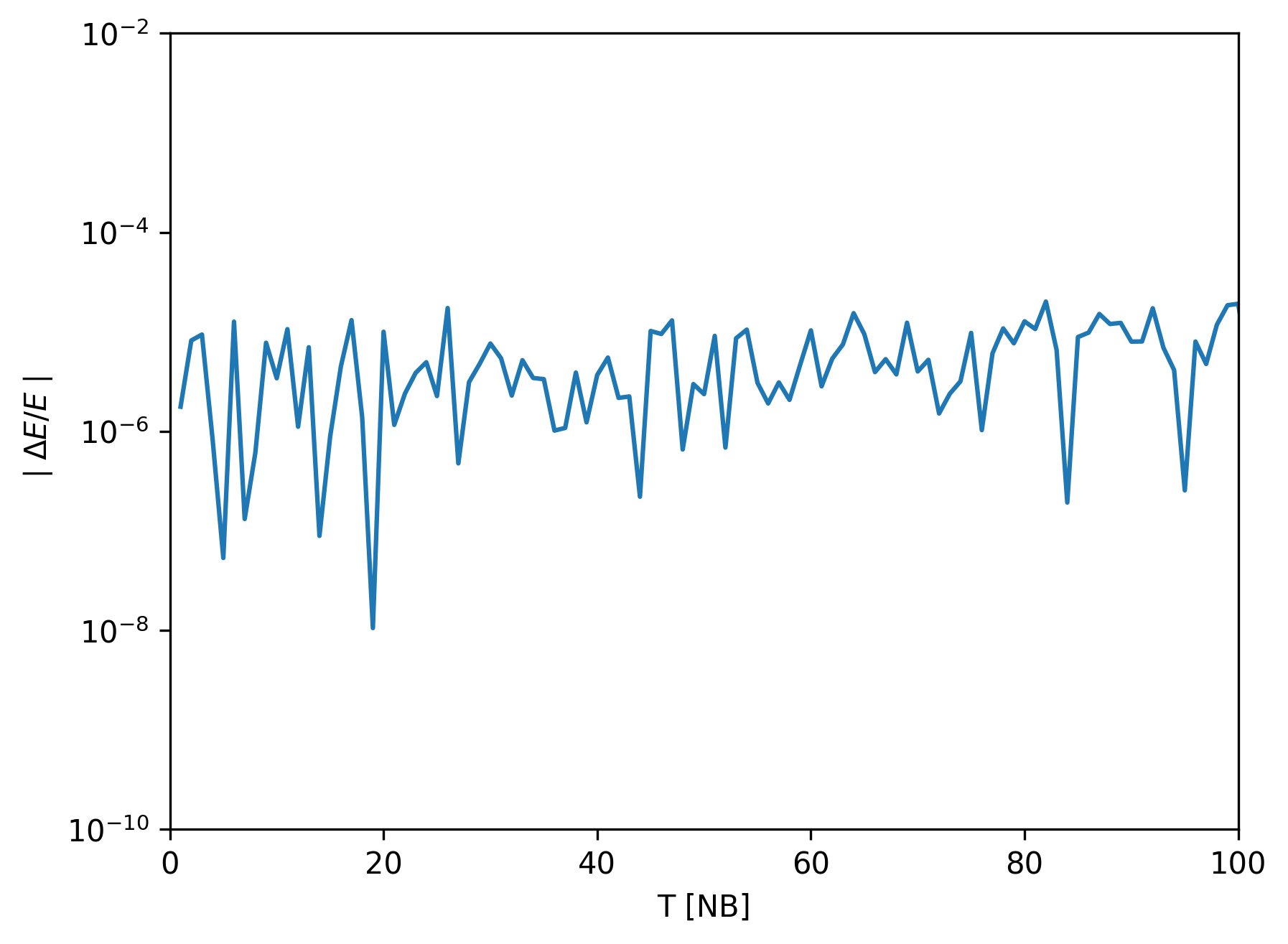} 
    \caption{Total relative energy error against time in N-body units of model A.}
    \label{energy}
\end{figure}

\begin{figure}
    \centering
    \includegraphics[width=\linewidth]{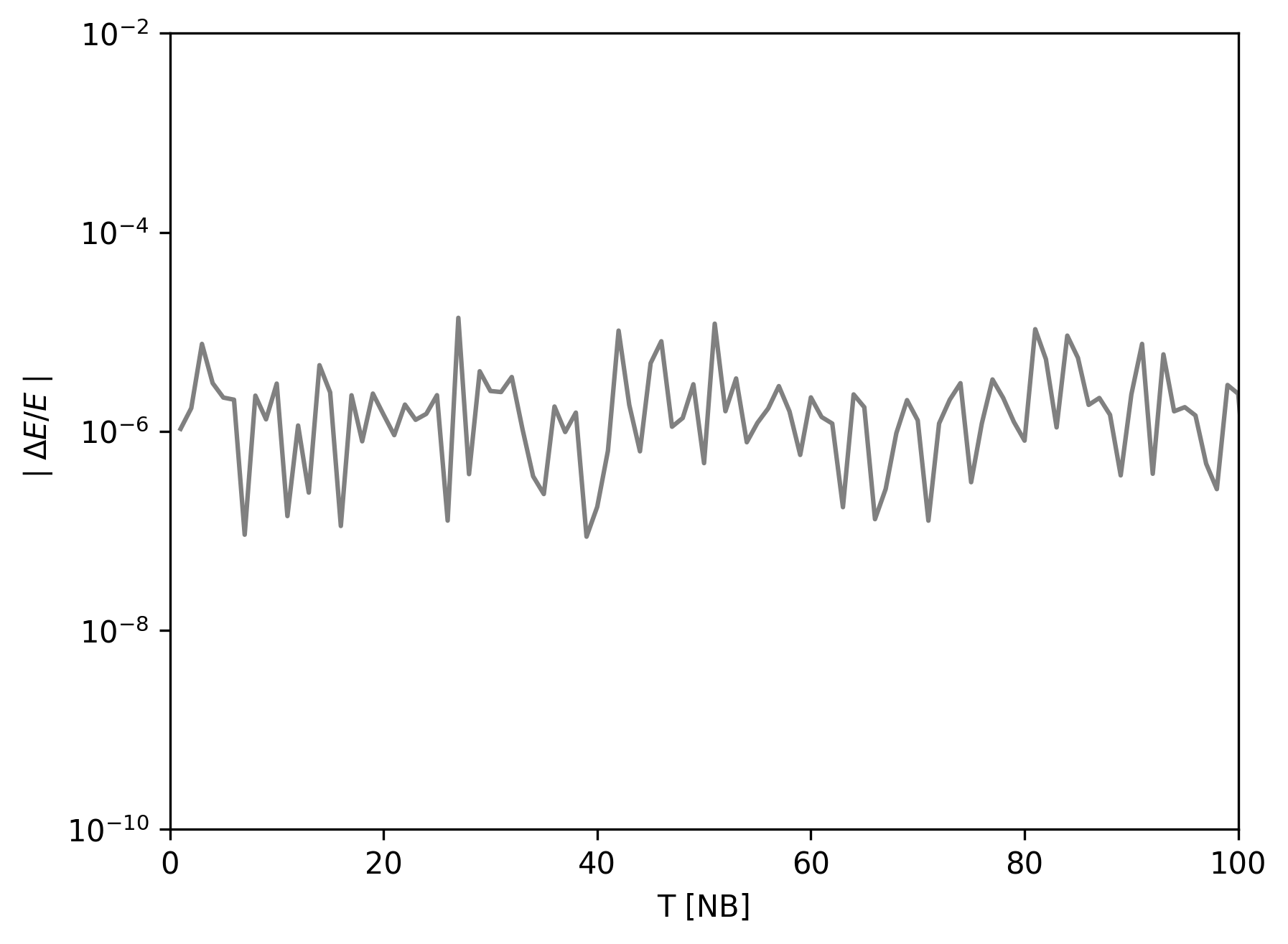} 
    \caption{Total relative energy error against time in N-body units of model D.}
    \label{energy2}
\end{figure}

\begin{figure}
    \centering
    \includegraphics[width=\linewidth]{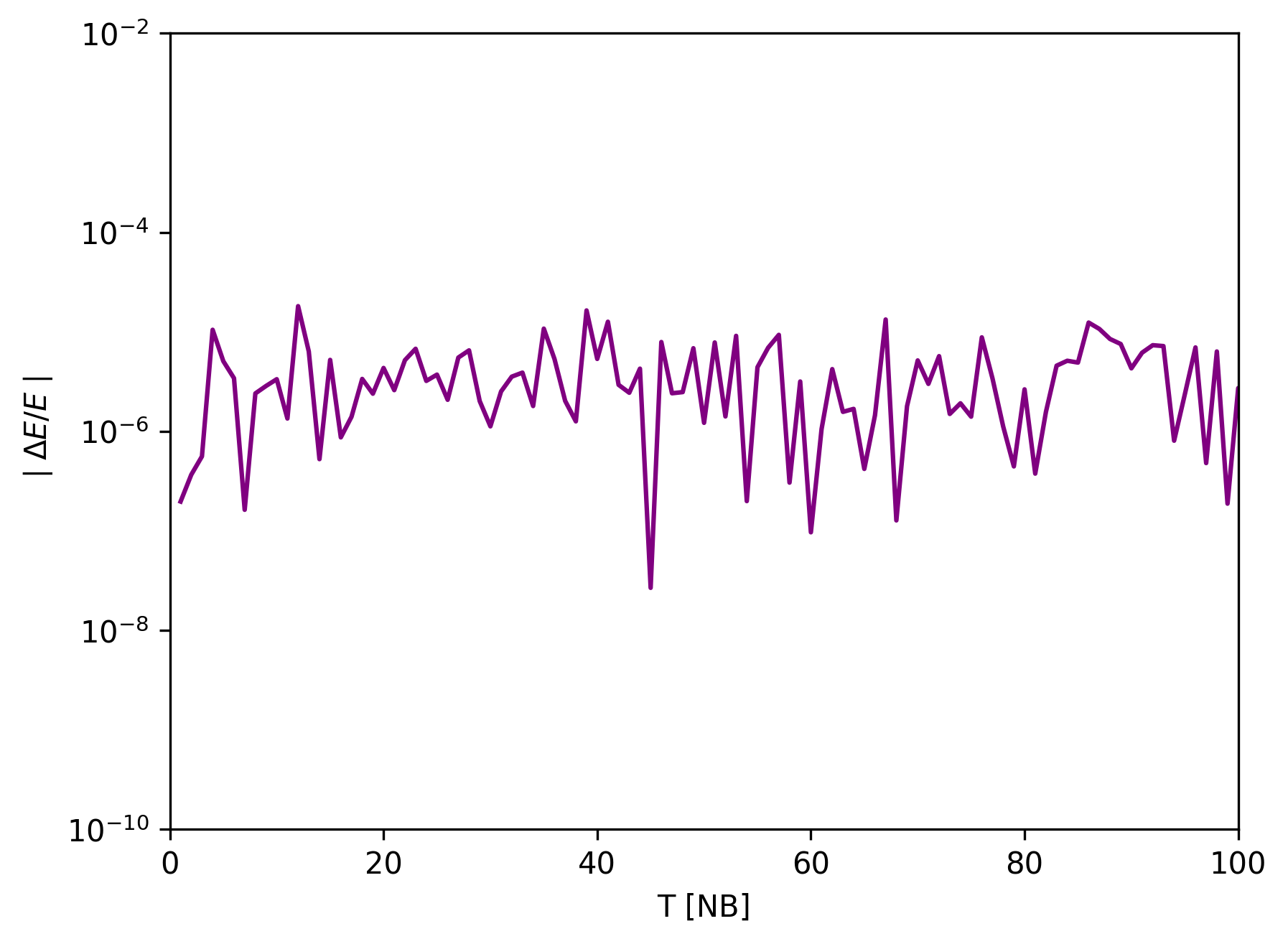} 
    \caption{Total relative energy error against time in N-body units of model I.}
    \label{energy3}
\end{figure}

In Figs.~\ref{energy}, \ref{energy2}, \ref{energy3} we show the total relative energy error against N-body time of models A, D, and I. The energy error of order $0.0001\%$ is such a small error that indicates that the total energy of the simulated systems has been conserved to a high degree of accuracy over the course of the simulations.

\bsp	
\label{lastpage}
\end{document}